\documentclass[article,shortnames]{jss_edited}

\usepackage{thumbpdf,lmodern}

\usepackage{framed}

\usepackage{amsmath,amssymb,amsthm,geometry,hyperref}
\usepackage{graphicx}						
\usepackage[utf8]{inputenc}
\usepackage[T1]{fontenc}
\usepackage{caption}
\usepackage{subcaption}

\usepackage[longtable]{multirow}
\usepackage{longtable}
\usepackage{multirow}
\usepackage{placeins}

\usepackage{float}

\let\Oldsection\section
\renewcommand{\section}{\FloatBarrier\Oldsection}

\let\Oldsubsection\subsection
\renewcommand{\subsection}{\FloatBarrier\Oldsubsection}

\let\Oldsubsubsection\subsubsection
\renewcommand{\subsubsection}{\FloatBarrier\Oldsubsubsection}

\newcommand{\be}{\begin{equation}}
\newcommand{\ee}{\end{equation}}
\newcommand{\bea}{\begin{eqnarray}}
\newcommand{\eea}{\end{eqnarray}}
\newcommand{\bean}{\begin{eqnarray*}}
	\newcommand{\eean}{\end{eqnarray*}}

\newcommand{\si}{\sigma}

\author{Ranjiva Munasinghe\\MIND Analytics \& Management 
	\And Pathum Kossinna\\University of Calgary
	\AND Dovini Jayasinghe\\MIND Analytics \& Management 
	\And Dilanka Wijeratne\\RMIT University}
\Plainauthor{Ranjiva Munasinghe, Pathum Kossinna, Dovini Jayasinghe, Dilanka Wijeratne}

\title{Fast Tail Index Estimation for Power Law Distributions in \proglang{R}}
\Plaintitle{Fast Tail Index Estimation for Power Law Distributions in R}
\Shorttitle{Fast Tail Index Estimation for Power Laws}

\Abstract{
	Power law distributions, in particular Pareto distributions, describe data across diverse areas of study. 
	We have developed a package in \proglang{R} to estimate the tail index for such datasets focusing on speed (in particular with large datasets), keeping in mind ease of use, as well as accuracy. In this short article, we provide a user guide to our package along with the results obtained highlighting the speed advantages of our package.
}

\Plainkeywords{Power law distributions, fat tailed distributions, heavy tailed distributions, Pareto distributions, tail index, tail estimation, fast computation, R}

\Address{
	Ranjiva Munasinghe\\
	CEO\\
	MIND Analytics \&  Management (Pvt) Ltd\\
	10/1 De Fonseka Place\\
	Colombo 5\\
	Sri Lanka\\
	E-mail: \email{ranjiva@mindlanka.org}\\
	URL: \url{www.mind-am.org}
}

\begin{document}	
	\section{Introduction} \label{sec:intro}
	Power law distributions account for phenomena and observations across diverse fields including physics, astronomy, biology, economics and the social sciences \citep{Sornette}. \cite{Newman} cites plenty of specific real-world examples including distribution of wealth, word frequency in a given text, population in a city etc. \cite{Taleb} states the most general form of the survival function\footnote{The cumulative density function (cdf) is $P(x)$, which is the probability that random variable $X$ is less than $x$. The complementary cumulative density function (ccdf) also referred to as the survival function is given by $\overline{P}(x) = 1 - P(x)$} describing power law distributions as
	\be \label{eq:PL_Gen}
	\overline{P}(x) = L(x) x^{-\alpha}
	\ee
	where $L(x)$ is a slowly-varying function, i.e.
	\[
	\lim_{x \rightarrow \infty} \frac{L(\lambda x)}{L(x)} = 1 \; \; \forall \lambda > 0
	\]
	The (tail) exponent $\alpha$ is referred to as the tail index\footnote{The tail index is also referred to as shape parameter in the context of Pareto distributions, stability parameter in the context of stable distributions and degrees of freedom in the context of t-distributions.}. The main motivation for this article and the accompanying \proglang{R} \citep{baseR} package is a study in a forthcoming publication \citep{RMM} analyzing power law data, in particular that of Pareto distributions, a special class of power laws with probability density function\footnote{The pdf is the derivative of the cdf: $p(x) = \frac{dP(x)}{dx}$} (pdf) defined for $x \geq x_{\mathrm{min}}>0$:
	\be \label{eq:Pareto1}
	p(x) = \frac{ \alpha x^{\alpha}_{\mathrm{min}}}{x^{1+\alpha}}
	\ee
	In the present \lq Big Data\rq era, it is essential for programs to be optimized for speed and have the ability to handle large data sets. In particular, there are applications in genetics involving large data sets where power-law distributions describe network connectivity \cite{BinZ}.  We found some of the existing tail estimation packages are not optimized for speed especially when dealing with a large data set. Our team then decided that it would be better to code our own parameter estimation tools in \proglang{R} with the main goal being fast computation of  the tail index $\alpha$. We would also like to have the ability to rapidly generate (large) Pareto data sets . This resulted in the creation of our \proglang{R}-package  \pkg{ptsuite}. 
	
	Our main objective with our estimator package is to address the speed advantage. In particular, we want to examine how long the code takes to run. The effect of the sample size on the speed was examined. We compared our run times against some existing \proglang{R} packages.
	The other objective is to test for the accuracy of our tail index estimates by seeing how close the tail index estimates are on known tail index values corresponding to generated Pareto data. As the tail index estimation algorithms are well-known, this is a secondary objective\footnote{In the Appendix \ref{sec:GPDevir}, we also look at the accuracy of the estimates produced by the other packages. All packages studied in this paper produce high accuracy estimates which improve with sample size.} We will also look at the size of the sample and the effect it has on accuracy. Finally, we look at tail index estimation for heavy tailed non-Pareto distributions.
	
	This rest of the paper is structured in the following format. We review the Tail Index Estimation methods for the functions used in the package \pkg{ptsuite} in Section \ref{sec:overview}. Next in Section \ref{sec:speedresults} we present our speed results by comparing run times for functions in \pkg{ptsuite} with their counterpart functions from other \proglang{R}-packages. We conclude this section with a brief comparison of run times of the functions in \pkg{ptsuite}. The next Section \ref{sec:accuresults} looks at the accuracy of the \pkg{ptsuite} tail index estimates, in particular focusing on the Hill Estimate. Section \ref{sec:usingthepackage} follows and is a guide on using the \pkg{ptsuite} package. We conclude with Section \ref{sec:summary} which summarizes this work and discusses future directions.
	
	\begin{leftbar}
		In this article our main considerations are Pareto distributions which are pure power law distributions as per the definition in equation (\ref{eq:Pareto1}). With regards to the general definition of power law distributions as per (\ref{eq:PL_Gen}) we see there are three scenarios:
		\begin{enumerate}
			\item Right tails, i.e. power law behavior in the limit $x \rightarrow \infty$. The Pareto family is in this class.
			\item Left tails, i.e. power law behavior in the limit $x \rightarrow - \infty$
			\item Both left and right tails, i.e. power law behavior in limit $x \rightarrow \pm \infty$
		\end{enumerate}
		Symmetric stable distributions and Student-T distributions fall into the last class. More general forms of stable distributions include a skew parameter that can result in exclusively left or right tails - for more details please refer to \cite{Nolan}.
	\end{leftbar}
	
	\newpage
	\section{Overview of Tail Index Estimation Methods} \label{sec:overview}
	To identify Pareto behavior we can start with a frequency plot/histogram and try to visually identify a lower bound (c.f. $x_{\mathrm{min}}$) and `strict power law' behavior. We may also do further heuristics such as log-log plots and log-log rank plots and look for a linear relationship to identify power law behavior, though these methods are not fool proof \cite{Nair}. In \cite{Hubert} there is another Pareto test given which looks for linear behavior in the QQ-plots of the log transformed data against standard exponential data\footnote{The slope of this line would give the $\alpha$ estimate.}. A simple implementation of this method is included in our package. We focus on the tail index estimators given below.
	
	\begin{leftbar} 
		Estimation methods 1 - 5 estimate the tail index working under the assumption of Pareto distributed data.  To apply these methods for the general power law form (\ref{eq:PL_Gen}), we would look to identify where tail (power law) behavior starts which is not a precise or easy task \citep{Nair}.  For a comprehensive list and the background theory, the interested reader is referred to \cite{Hill, Nair, Pokorna, Hubert}. The reference by  \cite{Hundred} is a useful catalog of Pareto-tail index estimation techniques.
	\end{leftbar}
	
	\begin{enumerate}
		\item Maximum Likelihood Estimation (MLE)  \citep{Newman} \\
		The MLE formula leads to a biased estimator $\hat{\alpha}$:
		\be
		\label{eq:MLE}
		\hat{\alpha} = N \cdot \left[ \sum_{i=1}^N \frac{x_i}{\hat{x}_{\mathrm{min}}} \right]^{-1}
		\ee 
		Here $x_i$ represents the data point for $i=1, \ldots, N$. The minimum value, $x_{\mathrm{min}}$, is estimated from the data set and hence denoted $\hat{x}_{\mathrm{min}}$. As MLE leads to asymptotic normality for the estimator, the error bounds may be given by the standard deviation:
		\be
		\hat{\si} = \frac{\sqrt{n+1}}{n} \cdot \hat{\alpha} 
		\ee
		The estimate (\ref{eq:MLE}) can be converted to an unbiased version $\alpha^*$ as follows \cite{Rizzo}:
		\[
		\alpha^* = \frac{n-2}{n} \cdot \hat{\alpha}
		\]
		\item Least Squares Estimation (LS) \citep{Zaher, Nair} \\
		The first step in this method is to sort the data in increasing order. Next for each value $i$ (of $N$ data points) we calculate $y_i$ the number of points greater than the $i^{th}$ data point. In this method one seeks to minimize the sum of the squared errors between the rank plot and the logarithm of the ccdf.
		The estimator is given by
		\be
		\label{eq:LS1}
		\hat{\alpha} = \frac{  \sum_{i=1}^N  \left( \hat{y_i} -   \frac{1 }{N} \sum_{i=1}^N \hat{y_i} \right) \left( \ln x_i  - \frac{1}{N} \sum_{i=1}^N \ln x_i \right) }{ \sum_{i=1}^N \left( \ln x_i  - \frac{1}{N} \sum_{i=1}^N \ln x_i \right)^2}
		\ee 
		\item Weighted Least Squares Estimation (WLS) \citep{Nair} \\
		The formulation is the same as for the LS method, except that the sum of squared errors is weighted. The choice of weight $w_i$ is defined to be
		\be
		\label{eq:WLSweight}
		w_i = \left[ \ln \left( \frac{x_i}{\hat{x}_{\mathrm{min}}} \right) \right]^{-1}
		\ee
		For this choice of weight the WLS method is closely related to the MLE, in that they will converge in the large $N$-limit.
		The tail index estimate is then given by
		\be
		\label{eq:WLS1}
		\hat{\alpha} = \frac{ - \sum_{i=1}^N \ln \left( \hat{y_i} / N \right) }{ \sum_{i=1}^N \ln \left( x_i / \hat{x}_{\mathrm{min}} \right)}
		\ee
		If there are no ties in the sorted data we may directly write:
		\be
		\label{eq:WLS2}
		\hat{\alpha} = \frac{ - \sum_{i=1}^N \ln \left[ (N+1-i) / N \right] }{ \sum_{i=1}^N \ln \left( x_i / \hat{x}_{\mathrm{min}} \right)}
		\ee
		
		\item Percentile Method (PM) \citep{Bhatti} \\
		The PM estimator for the tail index is given by
		\be 
		\label{eq:PM}
		\hat{\alpha} = \frac{ \ln 3}{ \ln (P^*_{75}) - \ln (P^*_{25})}
		\ee
		Here $P^*_{q}$ is the $q^{th}$ percentile of the data set. We also test a modified version using the median which we refer to here as the Modified Percentile Method (MPM). This is given by
		\be 
		\label{eq:MPM}
		\hat{\alpha} = \frac{ \ln 2}{ \ln (P^*_{75}) - \ln (P^*_{50})}
		\ee
		The final modification we test makes use of the geometric mean and is written here in the shorthand Geometric Mean Percentile Method (GMPM):
		\be
		\label{eq:GMPM}
		\hat{\alpha} = \frac{ 1 - \ln 4}{ \frac{1}{N} \sum_{i=1}^N \ln x_i - \ln (P^*_{75}) }
		\ee
		There are further variations on PM estimation in \cite{Bhatti}.
		
		\item Method of Moments (MoM) \citep{Rytgaard}, \\
		The MoM estimator is derived by equating the sample mean to the theoretical mean of the Pareto distribution.
		\be 
		\label{eq:MoM}
		\hat{\alpha} = \frac{ \sum_{i=1}^N x_i   }{ \sum_{i=1}^N x_i  - N \cdot \hat{x}_{\mathrm{min}}} 
		\ee
		The major drawback with this method is that if $\alpha \leq 1$ the estimator will not converge to the true value - a Pareto distribution with $\alpha \leq 1$ has infinite mean. (For this estimator $\hat{\alpha}\longrightarrow 1$ when $\alpha \leq 1$ \citep{Vytaras}) It is always better to use this method in conjunction with another to check the validity of the estimate. 
		\item Hill Estimator (HE) \citep{Nair, Pokorna, Hill} \\
		The HE method is a type of MLE method \citep{Hill, Hubert, Hundred} which can be used for general power law setting (\ref{eq:PL_Gen}) and one would need to specify where the tail starts. In particular the HE formula requires the specification of the tuning parameter (starting point) $k$. Identifying where the tail starts is an art in itself and requires a thorough examination of the data. If one chooses the parameter $k$ to be too large the variance of the estimator increases; on the other hand if it is too low the bias increases \citep{Hundred}. The data first needs to be sorted in increasing order and then we may write:
		\be
		\label{eq:Hill}
		\hat{\alpha} (k) = k \cdot \left[  \sum_{j=1}^k \ln \frac{x_{N-j+1}}{x_{N-k}} \right]^{-1}
		\ee
	\end{enumerate}

	
	\section{Results for Speed} \label{sec:speedresults}
	In this section we look at the performance of functions in the \pkg{ptsuite} package against their (comparable) counterparts in other packages. All estimator functions in the \pkg{ptsuite} package have been coded in  \proglang{C++} to ensure the highest possible performance. We compare the speed of the functions available in the package using the package \pkg{microbenchmark} \citep{microbenchR}. We compare the following estimators/functions from the \pkg{ptsuite} package and their counterparts:
	
	\begin{longtable}{cll} 
		\hline
		\textbf{Estimator/function}                      & \textbf{Package}  & \textbf{Function in Package}              \\* 
		\hline
		\multirow{3}{*}{Hill's Estimator}       & \pkg{ptsuite}  & \code{alpha\_hills}                     \\*
		& \pkg{laeken}   & \code{thetaHill}                        \\*
		& \pkg{evir}     & \code{gpd} (set \code{nextremes} accordingly)  \\* 
		\hline
		\multirow{2}{*}{Method of Moments}      & \pkg{ptsuite}  & \code{alpha\_moment}                    \\*
		& \pkg{laeken}   & \code{thetaMoment}                      \\* 
		\hline
		\multirow{4}{*}{MLE}                    & \pkg{ptsuite}  & \code{alpha\_mle}                       \\*
		& \pkg{EnvStats} & \code{epareto} (set \code{method = `mle'})  \\*
		& \pkg{evir}     & \code{gpd} (set \code{nextremes} to the sample size)                             \\*
		& \pkg{DeMAND}   & \code{pareto.fit}                       \\* 
		\hline
		\multirow{2}{*}{Leaset Squares}         & \pkg{ptsuite}  & \code{alpha\_ls}                       \\*
		& \pkg{EnvStats} & \code{epareto} (set \code{method = `ls'})   \\* 
		\hline
		\multirow{3}{*}{Pareto Data Generation} & \pkg{ptsuite}  & \code{generate\_pareto}                 \\*
		& \pkg{EnvStats} & \code{rpareto}                          \\*
		& \pkg{evir}     & \code{rgpd}                             \\
		\hline
		\caption{Existing \pkg{R} packages and their functions corresponding to functions in \pkg{ptsuite}.}
	\end{longtable}
	
	We complete a hundred runs for a selected sample using \pkg{ptsuite} and the counterpart. We capture the following times: mean, maximum, median and minimum. These times are all measured in microseconds ($\mu$s)\footnote{Times are displayed in log microseconds in the graphs. Sample sizes are presented in log units. Overall the display is better with these unit adjustments.}. In all the tests, \pkg{ptsuite} comes out on top. We should note however for some of the functions, the fastest (minimum) time of the competing package has been faster than the slowest (maximum) time of the \pkg{ptsuite}.  
	
	\begin{figure}[H]
		\begin{subfigure}{\textwidth}
			\centering \includegraphics{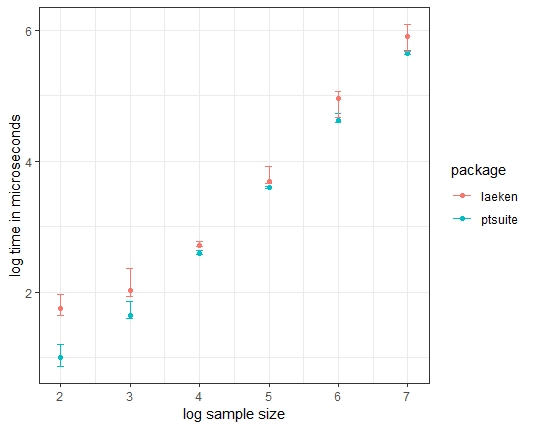} 
			\label{fig:hills1}
			\caption{Speed comparison of \pkg{ptsuite} and \pkg{laeken}.}
		\end{subfigure}
		\begin{subfigure}{\textwidth}
			\centering \includegraphics{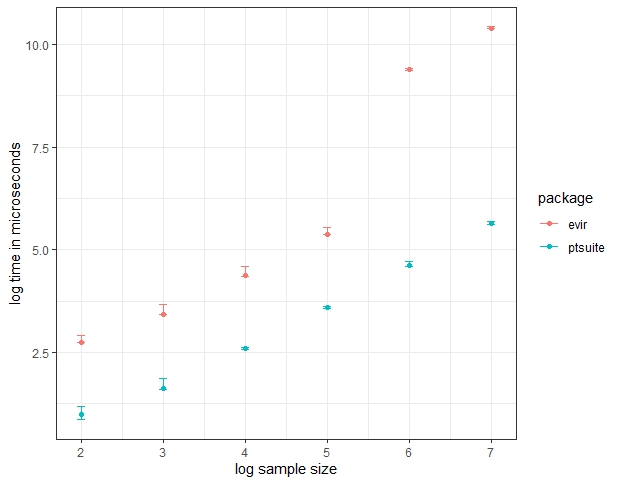} 
			\caption{Speed comparison of \pkg{ptsuite} and \pkg{evir}.}
			\label{fig:hills1_evir}
		\end{subfigure}
		\caption{Time plot of Hill's estimates for sample 1.}
	\end{figure}
	
	\small
	\begin{longtable}{clcccc} 
		\multicolumn{6}{r}{\tablename\ \thetable\ -- \textit{Continued on next page}} \\
		\endfoot
		\endlastfoot
		\hline
		\textbf{Sample Size}               & \textbf{Package} & \multicolumn{1}{l}{\textbf{Mean ($\mu$s)}} & \multicolumn{1}{l}{\textbf{Maximum ($\mu$s)}} & \multicolumn{1}{l}{\textbf{Median ($\mu$s)}} & \multicolumn{1}{l}{\textbf{Minimum ($\mu$s)}} \\* 
		\hline
		\multirow{3}{*}{$10^{2}$}      & ptsuite & 1.03E+01                        & 1.58E+01                       & 1.01E+01                          & 7.30E+00                        \\*
		& laeken  & 5.64E+01                        & 9.20E+01                       & 5.54E+01                          & 4.39E+01                        \\*
		& evir    & 5.77E+02                        & 8.18E+02                       & 5.66E+02                          & 5.59E+02                        \\* 
		\hline
		\multirow{3}{*}{$10^{3}$}     & ptsuite & 4.50E+01                        & 7.27E+01                       & 4.33E+01                          & 3.92E+01                        \\*
		& laeken  & 1.05E+02                        & 2.26E+02                       & 1.04E+02                          & 8.52E+01                        \\*
		& evir    & 2.81E+03                        & 4.69E+03                       & 2.67E+03                          & 2.61E+03                        \\* 
		\hline
		\multirow{3}{*}{$10^{4}$}    & ptsuite & 3.91E+02                        & 4.26E+02                       & 3.91E+02                          & 3.74E+02                        \\*
		& laeken  & 5.28E+02                        & 5.87E+02                       & 5.20E+02                          & 4.96E+02                        \\*
		& evir    & 2.44E+04                        & 3.92E+04                       & 2.36E+04                          & 2.29E+04                        \\* 
		\hline
		\multirow{3}{*}{$10^{5}$}   & ptsuite & 3.94E+03                        & 4.17E+03                       & 3.97E+03                          & 3.79E+03                        \\*
		& laeken  & 5.92E+03                        & 8.23E+03                       & 4.82E+03                          & 4.51E+03                        \\*
		& evir    & 2.40E+05                        & 3.47E+05                       & 2.38E+05                          & 2.32E+05                        \\* 
		\hline
		\multirow{3}{*}{$10^{6}$}  & ptsuite & 4.18E+04                        & 5.28E+04                       & 4.19E+04                          & 3.95E+04                        \\*
		& laeken  & 7.21E+04                        & 1.15E+05                       & 9.19E+04                          & 4.67E+04                        \\*
		& evir    & 2.47E+09                        & 2.58E+09                       & 2.49E+09                          & 2.37E+09                        \\* 
		\hline
		\multirow{3}{*}{$10^{7}$} & ptsuite & 4.47E+05                        & 4.92E+05                       & 4.44E+05                          & 4.21E+05                        \\*
		& laeken  & 7.10E+05                        & 1.23E+06                       & 8.11E+05                          & 4.69E+05                        \\*
		& evir    & 2.49E+10                        & 2.67E+10                       & 2.46E+10                          & 2.41E+10                        \\
		\hline
		\caption{\label{tab:hills1} Function timing results using \pkg{microbenchmark} for Hill's Estimate with sample 1.}
	\end{longtable}
	
	\normalsize

	\begin{figure}[H]
		\centering \includegraphics{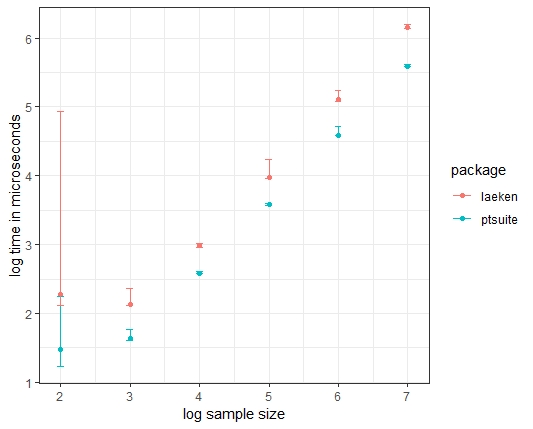} 
		\caption{Time plot of MoM estimates for sample 2.}
		\label{fig:mom2}
	\end{figure}
	
	\small
	\begin{longtable}{clcccc} 
		\multicolumn{6}{r}{\tablename\ \thetable\ -- \textit{Continued on next page}} \\
		\endfoot
		\endlastfoot
		\hline
		\textbf{Sample Size}               & \textbf{Package} & \multicolumn{1}{l}{\textbf{Mean ($\mu$s)}} & \multicolumn{1}{l}{\textbf{Maximum ($\mu$s)}} & \multicolumn{1}{l}{\textbf{Median ($\mu$s)}} & \multicolumn{1}{l}{\textbf{Minimum ($\mu$s)}} \\* 
		\hline
		\multirow{2}{*}{$10^{2}$} & ptsuite & 2.95E+01                   & 1.77E+02                      & 3.06E+01                     & 1.71E+01                       \\*
		& laeken  & 1.04E+03                   & 8.49E+04                      & 1.90E+02                     & 1.30E+02                       \\* 
		\hline
		\multirow{2}{*}{$10^{3}$} & ptsuite & 4.29E+01                   & 5.81E+01                      & 4.32E+01                     & 4.09E+01                       \\*
		& laeken  & 1.36E+02                   & 2.30E+02                      & 1.35E+02                     & 1.32E+02                       \\* 
		\hline
		\multirow{2}{*}{$10^{4}$} & ptsuite & 3.85E+02                   & 4.03E+02                      & 3.85E+02                     & 3.80E+02                       \\*
		& laeken  & 9.60E+02                   & 1.03E+03                      & 9.74E+02                     & 9.06E+02                       \\* 
		\hline
		\multirow{2}{*}{$10^{5}$} & ptsuite & 3.79E+03                   & 3.92E+03                      & 3.79E+03                     & 3.76E+03                       \\*
		& laeken  & 9.81E+03                   & 1.72E+04                      & 9.36E+03                     & 9.25E+03                       \\* 
		\hline
		\multirow{2}{*}{$10^{6}$} & ptsuite & 3.91E+04                   & 5.12E+04                      & 3.87E+04                     & 3.81E+04                       \\*
		& laeken  & 1.33E+05                   & 1.71E+05                      & 1.29E+05                     & 1.19E+05                       \\* 
		\hline
		\multirow{2}{*}{$10^{7}$} & ptsuite & 3.86E+05                   & 4.12E+05                      & 3.84E+05                     & 3.82E+05                       \\*
		& laeken  & 1.44E+06                   & 1.58E+06                      & 1.43E+06                     & 1.39E+06                       \\
		\hline
		\caption{\label{tab:mom2} Function timing results using \pkg{microbenchmark} for MoM Estimate with sample 2.}
	\end{longtable}
	\normalsize
	
	\newpage
	\begin{figure}[H]
		\begin{subfigure}{\textwidth}
			\centering \includegraphics{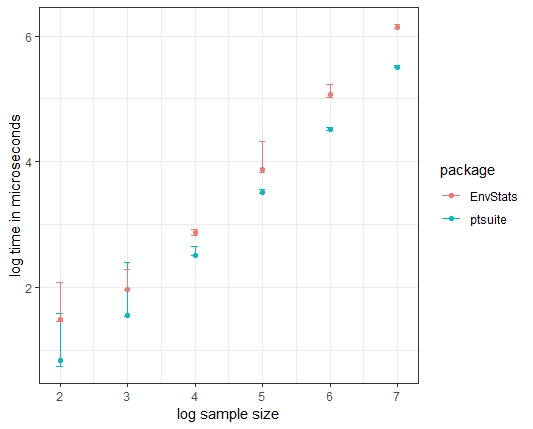} 
			\caption{Speed comparison of \pkg{ptsuite} and \pkg{EnvStats}.}
			\label{fig:MLE3}
		\end{subfigure}
		\begin{subfigure}{\textwidth}
			\centering \includegraphics{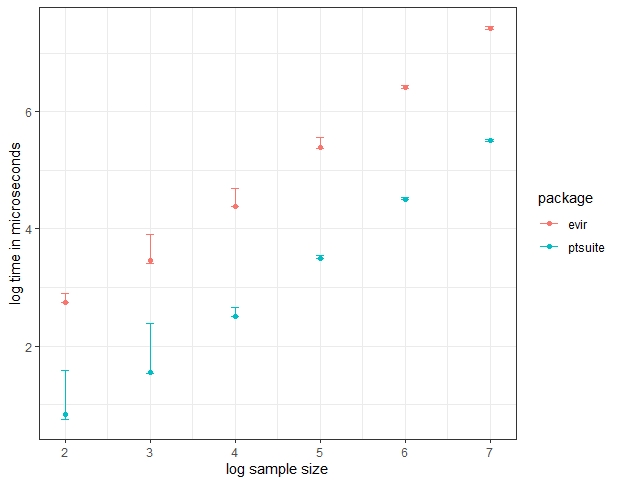} 
			\caption{Speed comparison of \pkg{ptsuite} and \pkg{evir}.}
			\label{fig:MLE3_evir}
		\end{subfigure}
		\caption{Time plots of MLE estimates for sample 3.}
	\end{figure}
	
	\small
	\begin{longtable}{clcccc} 
		\hline
		\textbf{Sample Size}               & \textbf{Package} & \multicolumn{1}{l}{\textbf{Mean ($\mu$s)}} & \multicolumn{1}{l}{\textbf{Maximum ($\mu$s)}} & \multicolumn{1}{l}{\textbf{Median ($\mu$s)}} & \multicolumn{1}{l}{\textbf{Minimum ($\mu$s)}} \\* 
		\hline
		\multirow{3}{*}{$10^{2}$}      & ptsuite  & 7.07E+00                      & 3.83E+01                         & 6.80E+00                        & 5.50E+00                          \\*
		& EnvStats & 3.23E+01                      & 1.20E+02                         & 3.03E+01                        & 2.85E+01                          \\*
		& evir     & 5.72E+02                      & 7.94E+02                         & 5.66E+02                        & 5.59E+02                          \\* 
		\hline
		\multirow{3}{*}{$10^{3}$}     & ptsuite  & 3.91E+01                      & 2.47E+02                         & 3.58E+01                        & 3.41E+01                          \\*
		& EnvStats & 9.80E+01                      & 1.93E+02                         & 9.15E+01                        & 8.73E+01                          \\*
		& evir     & 2.94E+03                      & 8.18E+03                         & 2.93E+03                        & 2.61E+03                          \\* 
		\hline
		\multirow{3}{*}{$10^{4}$}    & ptsuite  & 3.23E+02                      & 4.50E+02                         & 3.21E+02                        & 3.17E+02                          \\*
		& EnvStats & 7.38E+02                      & 8.19E+02                         & 7.35E+02                        & 6.60E+02                          \\*
		& evir     & 2.57E+04                      & 4.87E+04                         & 2.46E+04                        & 2.41E+04                          \\* 
		\hline
		\multirow{3}{*}{$10^{5}$}   & ptsuite  & 3.22E+03                      & 3.54E+03                         & 3.19E+03                        & 3.15E+03                          \\*
		& EnvStats & 7.94E+03                      & 2.10E+04                         & 7.53E+03                        & 6.84E+03                          \\*
		& evir     & 2.47E+05                      & 3.70E+05                         & 2.45E+05                        & 2.33E+05                          \\* 
		\hline
		\multirow{3}{*}{$10^{6}$}  & ptsuite  & 3.23E+04                      & 3.48E+04                         & 3.21E+04                        & 3.18E+04                          \\*
		& EnvStats & 1.23E+05                      & 1.70E+05                         & 1.18E+05                        & 1.06E+05                          \\*
		& evir     & 2.66E+06                      & 2.84E+06                         & 2.66E+06                        & 2.54E+06                          \\* 
		\hline
		\multirow{3}{*}{$10^{7}$} & ptsuite  & 3.21E+05                      & 3.35E+05                         & 3.20E+05                        & 3.17E+05                          \\*
		& EnvStats & 1.40E+06                      & 1.53E+06                         & 1.39E+06                        & 1.30E+06                          \\*
		& evir     & 2.66E+07                      & 2.83E+07                         & 2.65E+07                        & 2.61E+07                          \\
		\hline
		\caption{\label{tab:mle3} Function timing results using \pkg{microbenchmark} for MLE Estimate with sample 3.}
	\end{longtable}
	\normalsize
	
	\newpage
	\begin{figure}[H]
		\centering \includegraphics{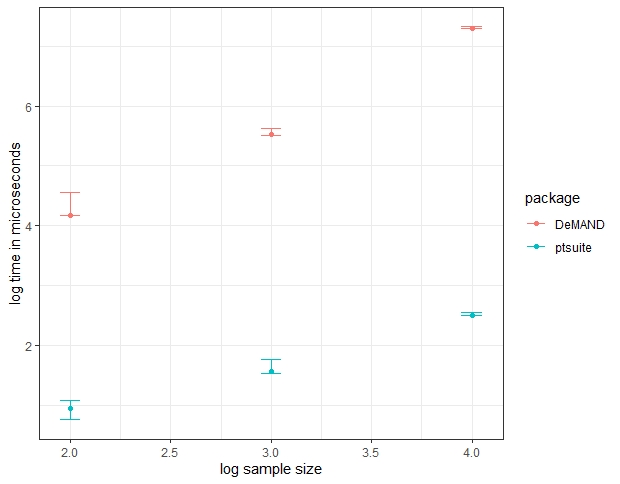} 
		\caption{Time plot of MLE estimates for sample 1.}
		\label{fig:mle1_demand}
	\end{figure}
	
	\small
	\begin{longtable}{clcccc} 
		\hline
		\textbf{Sample Size}               & \textbf{Package} & \multicolumn{1}{l}{\textbf{Mean ($\mu$s)}} & \multicolumn{1}{l}{\textbf{Maximum ($\mu$s)}} & \multicolumn{1}{l}{\textbf{Median ($\mu$s)}} & \multicolumn{1}{l}{\textbf{Minimum ($\mu$s)}} \\* 
		\hline
		\multirow{2}{*}{$10^{2}$}   & ptsuite & 8.61E+00                        & 1.21E+01                       & 8.80E+00                          & 5.80E+00                        \\*
		& DeMAND  & 1.59E+04                        & 3.64E+04                       & 1.50E+04                          & 1.47E+04                        \\* 
		\hline
		\multirow{2}{*}{$10^{3}$}  & ptsuite & 3.90E+01                        & 5.78E+01                       & 3.77E+01                          & 3.44E+01                        \\*
		& DeMAND  & 3.33E+05                        & 4.14E+05                       & 3.29E+05                          & 3.16E+05                        \\* 
		\hline
		\multirow{2}{*}{$10^{4}$} & ptsuite & 3.27E+02                        & 3.52E+02                       & 3.24E+02                          & 3.18E+02                        \\*
		& DeMAND  & 1.99E+07                        & 2.09E+07                       & 1.99E+07                          & 1.95E+07                        \\
		\hline
		\caption{\label{tab:mle1_demand} Function timing results using \pkg{microbenchmark} for MLE Estimate with sample 1.}
	\end{longtable}
	\normalsize
	
	\begin{figure}[H]
		\centering \includegraphics{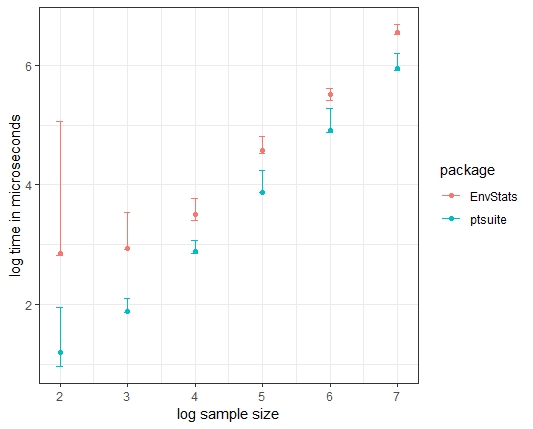} 
		\caption{Time plot of LS estimates for sample 4.}
		\label{fig:ls4}
	\end{figure}
	
	\small
	\begin{longtable}{clcccc} 
		\hline
		\textbf{Sample Size}               & \textbf{Package} & \multicolumn{1}{l}{\textbf{Mean ($\mu$s)}} & \multicolumn{1}{l}{\textbf{Maximum ($\mu$s)}} & \multicolumn{1}{l}{\textbf{Median ($\mu$s)}} & \multicolumn{1}{l}{\textbf{Minimum ($\mu$s)}} \\* 
		\hline
		\multirow{2}{*}{$10^{2}$}    & ptsuite & 2.02E+01                   & 8.80E+01                      & 1.55E+01                     & 9.20E+00                       \\*
		& EnvStats  & 2.10E+03                   & 1.17E+05                      & 7.23E+02                     & 6.65E+02                       \\* 
		\hline
		\multirow{2}{*}{$10^{3}$}     & ptsuite & 7.83E+01                   & 1.28E+02                      & 7.73E+01                     & 7.28E+01                       \\*
		& EnvStats  & 9.06E+02                   & 3.47E+03                      & 8.65E+02                     & 8.19E+02                       \\* 
		\hline
		\multirow{2}{*}{$10^{4}$}    & ptsuite & 7.81E+02                   & 1.20E+03                      & 7.73E+02                     & 7.26E+02                       \\*
		& EnvStats  & 3.19E+03                   & 5.84E+03                      & 3.20E+03                     & 2.52E+03                       \\* 
		\hline
		\multirow{2}{*}{$10^{5}$}   & ptsuite & 7.84E+03                   & 1.73E+04                      & 7.56E+03                     & 7.34E+03                       \\*
		& EnvStats  & 4.12E+04                   & 6.60E+04                      & 3.77E+04                     & 3.42E+04                       \\* 
		\hline
		\multirow{2}{*}{$10^{6}$}  & ptsuite & 9.45E+04                   & 1.91E+05                      & 8.06E+04                     & 7.66E+04                       \\*
		& EnvStats  & 3.28E+05                   & 4.15E+05                      & 3.23E+05                     & 2.63E+05                       \\* 
		\hline
		\multirow{2}{*}{$10^{7}$} & ptsuite & 9.17E+05                   & 1.61E+06                      & 8.79E+05                     & 8.09E+05                       \\*
		& EnvStats  & 3.66E+06                   & 4.87E+06                      & 3.62E+06                     & 3.34E+06                       \\
		\hline
		\caption{\label{tab:ls4} Function timing results using \pkg{microbenchmark} for LS Estimate with sample 4.}
	\end{longtable}
	\normalsize
	\newpage
	\begin{figure}[H]
		\begin{subfigure}{\textwidth}
			\centering \includegraphics{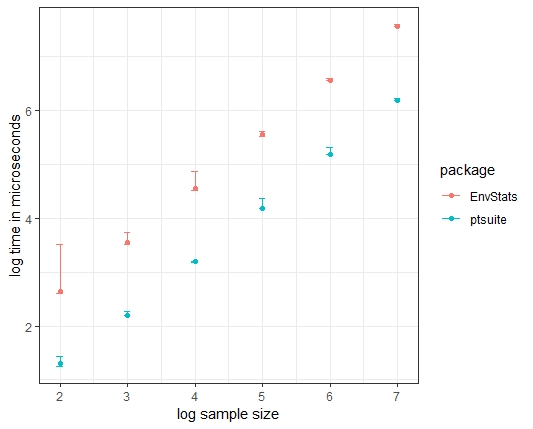} 
			\caption{Speed comparison of \pkg{ptsuite} and \pkg{EnvStats}.}
			\label{fig:gp5}
		\end{subfigure}
		\begin{subfigure}{\textwidth}
			\centering \includegraphics{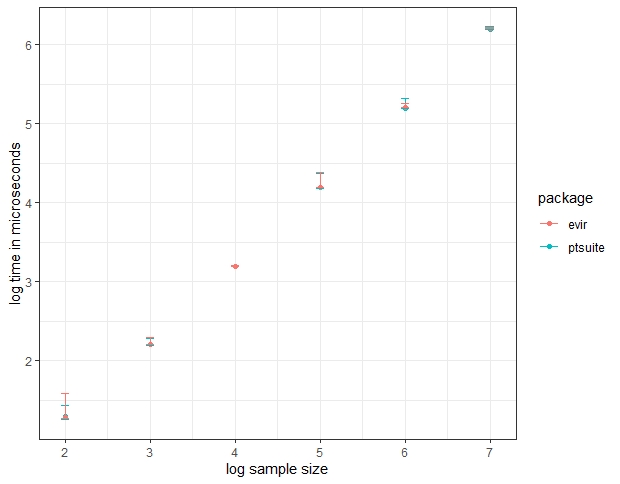} 
			\caption{Speed comparison of \pkg{ptsuite} and \pkg{evir}.}
			\label{fig:gp5_evir}
		\end{subfigure}
		\caption{Time plots of generating Pareto data for sample 5.}
	\end{figure}
	
	\small
	\begin{longtable}{clcccc} 
		\hline
		\textbf{Sample Size}               & \textbf{Package} & \multicolumn{1}{l}{\textbf{Mean ($\mu$s)}} & \multicolumn{1}{l}{\textbf{Maximum ($\mu$s)}} & \multicolumn{1}{l}{\textbf{Median ($\mu$s)}} & \multicolumn{1}{l}{\textbf{Minimum ($\mu$s)}} \\* 
		\hline
		\multirow{3}{*}{$10^{2}$}      & ptsuite  & 2.02E+01                        & 2.71E+01                       & 2.00E+01                          & 1.79E+01                        \\*
		& EnvStats & 4.89E+02                        & 3.23E+03                       & 4.48E+02                          & 4.07E+02                        \\*
		& evir     & 1.97E+01                        & 3.87E+01                       & 1.93E+01                          & 1.86E+01                        \\* 
		\hline
		\multirow{3}{*}{$10^{3}$}     & ptsuite  & 1.60E+02                        & 1.91E+02                       & 1.59E+02                          & 1.56E+02                        \\*
		& EnvStats & 3.63E+03                        & 5.44E+03                       & 3.51E+03                          & 3.24E+03                        \\*
		& evir     & 1.62E+02                        & 1.96E+02                       & 1.61E+02                          & 1.59E+02                        \\* 
		\hline
		\multirow{3}{*}{$10^{4}$}    & ptsuite  & 1.55E+03                        & 1.60E+03                       & 1.55E+03                          & 1.54E+03                        \\*
		& EnvStats & 3.58E+04                        & 7.29E+04                       & 3.52E+04                          & 3.23E+04                        \\*
		& evir     & 1.57E+03                        & 1.62E+03                       & 1.57E+03                          & 1.56E+03                        \\* 
		\hline
		\multirow{3}{*}{$10^{5}$}   & ptsuite  & 1.61E+04                        & 2.33E+04                       & 1.56E+04                          & 1.54E+04                        \\*
		& EnvStats & 3.62E+05                        & 4.05E+05                       & 3.58E+05                          & 3.31E+05                        \\*
		& evir     & 1.57E+04                        & 2.42E+04                       & 1.56E+04                          & 1.56E+04                        \\* 
		\hline
		\multirow{3}{*}{$10^{6}$}  & ptsuite  & 1.59E+05                        & 2.11E+05                       & 1.56E+05                          & 1.55E+05                        \\*
		& EnvStats & 3.69E+06                        & 4.02E+06                       & 3.67E+06                          & 3.59E+06                        \\*
		& evir     & 1.61E+05                        & 1.79E+05                       & 1.60E+05                          & 1.58E+05                        \\* 
		\hline
		\multirow{3}{*}{$10^{7}$} & ptsuite  & 1.57E+06                        & 1.68E+06                       & 1.56E+06                          & 1.55E+06                        \\*
		& EnvStats & 3.67E+07                        & 4.00E+07                       & 3.64E+07                          & 3.59E+07                        \\*
		& evir     & 1.60E+06                        & 1.67E+06                       & 1.60E+06                          & 1.59E+06                        \\
		\hline
		\caption{\label{tab:gp5} Function timing results using \pkg{microbenchmark} for generating Pareto data with sample 5.}
	\end{longtable}
	\normalsize
	\subsection{Comparison of Estimation Methods within ptsuite package}
	In this section, we compare and present the speed of the (estimation method) functions available in the package \pkg{ptsuite} using the package \pkg{microbenchmark} \citep{microbenchR}. We do not include Hill's Estimator as it restricts the domain to a subset of the sample. The results of this comparison can be seen in Table \ref{tab:speedptsuite}.
	
	We note here that the most computationally intensive functions are the LS and the WLS methods with the WLS method consuming a very large time for progressively larger datasets. It is seen that the PM and the MPM are consistently the fastest methods regardless of the sample size (but we remind the reader that the results of the percentile methods are also the least accurate for small sample sizes).
	
		\begin{figure}[H]
		\centering
		\includegraphics{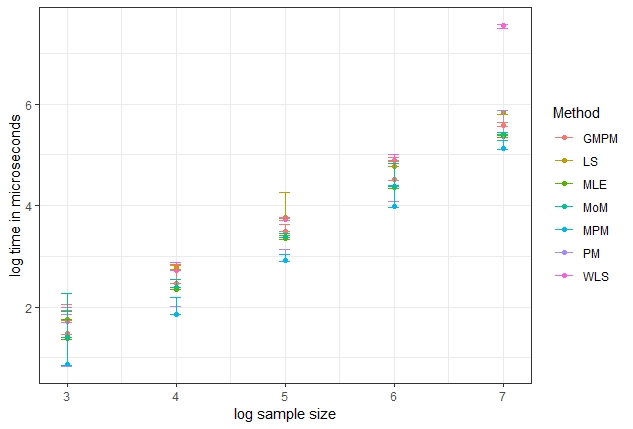} 
		\caption{Function timing results using \pkg{microbenchmark} for various sample sizes.}
		\label{fig:all_ptsuite_speeds}
	\end{figure}

	\small
	\begin{longtable}{lcrrr} 
		\multicolumn{5}{r}{\tablename\ \thetable\ -- \textit{Continued on next page}} \\
		\endfoot
		\endlastfoot
		\hline
		\textbf{Method}  & \textbf{Sample Size}  & \textbf{Minimum ($\mu$s)}  & \textbf{Median ($\mu$s)}  & \textbf{Maximum ($\mu$s)}   \endfirsthead 
		\hline
		\multirow{5}{*}{MLE}  & $10^{3}$   & 2.36E+01                      & 2.45E+01                     & 5.77E+01                       \\*
		& $10^{4}$   & 2.20E+02                      & 2.23E+02                     & 3.55E+02                       \\*
		& $10^{5}$   & 2.17E+03                      & 2.23E+03                     & 2.63E+03                       \\*
		& $10^{6}$   & 2.19E+04                      & 2.23E+04                     & 2.33E+04                       \\*
		& $10^{7}$   & 2.22E+05                      & 2.30E+05                     & 2.50E+05                       \\* 
		\hline
		\multirow{5}{*}{LS}   & $10^{3}$   & 5.46E+01                      & 5.63E+01                     & 8.53E+01                       \\*
		& $10^{4}$   & 5.44E+02                      & 5.92E+02                     & 6.81E+02                       \\*
		& $10^{5}$   & 5.56E+03                      & 5.82E+03                     & 1.83E+04                       \\*
		& $10^{6}$   & 5.88E+04                      & 5.96E+04                     & 7.32E+04                       \\*
		& $10^{7}$   & 6.27E+05                      & 6.79E+05                     & 7.53E+05                       \\* 
		\hline
		\multirow{5}{*}{WLS}  & $10^{3}$   & 5.09E+01                      & 5.26E+01                     & 9.79E+01                       \\*
		& $10^{4}$   & 5.17E+02                      & 5.39E+02                     & 7.50E+02                       \\*
		& $10^{5}$   & 5.20E+03                      & 5.38E+03                     & 5.78E+03                       \\*
		& $10^{6}$   & 7.70E+04                      & 7.82E+04                     & 1.04E+05                       \\*
		& $10^{7}$   & 3.12E+07                      & 3.53E+07                     & 3.70E+07                       \\* 
		\hline
		\multirow{5}{*}{PM}   & $10^{3}$   & 6.83E+00                      & 7.40E+00                     & 7.14E+01                       \\*
		& $10^{4}$   & 7.08E+01                      & 7.20E+01                     & 1.04E+02                       \\*
		& $10^{5}$   & 8.01E+02                      & 8.17E+02                     & 1.36E+03                       \\*
		& $10^{6}$   & 9.37E+03                      & 9.66E+03                     & 1.18E+04                       \\*
		& $10^{7}$   & 1.30E+05                      & 1.36E+05                     & 7.45E+05                       \\* 
		\hline
		\multirow{5}{*}{MPM}  & $10^{3}$   & 7.11E+00                      & 7.54E+00                     & 8.16E+01                       \\*
		& $10^{4}$   & 7.05E+01                      & 7.23E+01                     & 1.52E+02                       \\*
		& $10^{5}$   & 8.02E+02                      & 8.15E+02                     & 1.11E+03                       \\*
		& $10^{6}$   & 9.35E+03                      & 9.66E+03                     & 2.44E+04                       \\*
		& $10^{7}$   & 1.30E+05                      & 1.36E+05                     & 1.88E+05                       \\* 
		\hline
		\multirow{5}{*}{GMPM} & $10^{3}$   & 2.90E+01                      & 3.02E+01                     & 1.12E+02                       \\*
		& $10^{4}$   & 2.93E+02                      & 2.95E+02                     & 6.61E+02                       \\*
		& $10^{5}$   & 3.03E+03                      & 3.07E+03                     & 4.19E+03                       \\*
		& $10^{6}$   & 3.19E+04                      & 3.24E+04                     & 8.90E+04                       \\*
		& $10^{7}$   & 3.55E+05                      & 3.72E+05                     & 4.34E+05                       \\* 
		\hline
		\multirow{5}{*}{MoM}  & $10^{3}$   & 2.56E+01                      & 2.70E+01                     & 1.83E+02                       \\*
		& $10^{4}$   & 2.39E+02                      & 2.44E+02                     & 3.44E+02                       \\*
		& $10^{5}$   & 2.36E+03                      & 2.42E+03                     & 2.81E+03                       \\*
		& $10^{6}$   & 2.40E+04                      & 2.43E+04                     & 6.68E+04                       \\*
		& $10^{7}$   & 2.42E+05                      & 2.52E+05                     & 2.79E+05                       \\
		\hline
		\caption{\label{tab:speedptsuite} Function timing results using \pkg{microbenchmark} for various sample sizes.}
	\end{longtable}
	\normalsize

	
	\section{Results for Accuracy} \label{sec:accuresults}
\begin{figure}[H]
		\centering
		\includegraphics{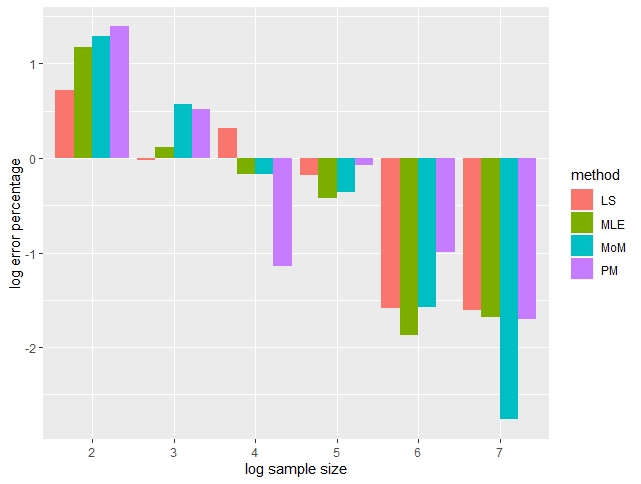}
		\caption{Accuracy plot of selected methods of estimation available in \pkg{ptsuite}. Sample results for $\alpha$ = 2.2 (using sample 1 from each run).}
		\label{fig:LS_MLE_MoM_PM_accuracy}
	\end{figure}

	We begin by noting that we test the HE method separately as we have to supply an extra parameter, namely the tuning parameter $k$. To test each of the other six methods methods we do the following:
	\begin{enumerate}
		\item Simulate Pareto data for 4 different values of $\alpha$: 0.5, 1.5, 2.2, 5. Pareto data is generated via
		\[
		x_i = \frac{x_{\mathrm{min}}}{u_i^{1/\alpha}}
		\]
		where $u_i$ is a standard uniform random number. We work with $x_{\mathrm{min}}=2$ for all the tests.
		\item For each  $\alpha$ value we generate 3 different sample sizes: $10^3, 10^5, 10^6$ to test the effect of sample size on the accuracy.
		\item For each sample size we generate four different samples (by changing the random number generator seed). 
	\end{enumerate} 
	In total we run $288 = 6 \times 4 \times 3 \times 4$ tests. To economize on space we present a sample of the results (Table \ref{tab:alpha2.2}) with the complete set of results given in Appendix \ref{sec:App-Est}. The $\hat{\alpha}$ error given is calculated as follows:
	\[
	\mathrm{Error \,\%} = \frac{|\alpha - \hat{\alpha}|}{\alpha} \times 100\%
	\]
	We also remark here that for the LS and PM methods, there is a peculiarity in that the accuracy dips slightly for particular increases in sample size. The reader is referred to Figure \ref{fig:LS_MLE_MoM_PM_accuracy} to observe the counter-intuitive behavior\footnote{Note that the error percentages are in log scale - hence larger negative bars indicate smaller error.}. For example when considering sample 1, the LS $\alpha$-estimate with $10^{3}$ samples has error $0.95\%$ and an error of $2.08\%$ with $10^{4}$ samples. When we increase the sample size to $10^{5}$, the error drops as expected. Similarly for sample 1, the PM $\alpha$-estimate with $10^{4}$ samples has error $0.07\%$ and an error of $0.85\%$ with $10^{5}$ samples. As with the LS method, when we increase the sample size to $10^{6}$, the error drops as expected. We are not certain as to the reason for the drop in accuracy for a certain values of larger samples - we consider other samples (by changing the seed) and these results are presented in Appendix \ref{sec:App-LS} and Appendix \ref{sec:App-PM} respectively. The other samples also have this counter-intuitive behavior but at different sample sizes.
	 
	\begin{table}[H]
		\small
		\centering
		\begin{tabular}{lcrr} 
			\hline
			\textbf{Method}               & \multicolumn{1}{l}{\textbf{Sample Size}} & \multicolumn{1}{l}{\textbf{$\hat{\alpha}$}} & \multicolumn{1}{l}{\textbf{$\hat{\alpha}$ error}}  \\* 
			\hline
			\multirow{6}{*}{MLE} & $10^{2}$   & 2.52705                                      & 14.87\%                                                                 \\*
			& $10^{3}$   & 2.22869                                      & 1.30\%                                                                  \\*
			& $10^{4}$   & 2.18519                                      & 0.67\%                                                                  \\*
			& $10^{5}$   & 2.19168                                      & 0.38\%                                                                  \\*
			& $10^{6}$   & 2.19971                                      & 0.01\%                                                                  \\*
			& $10^{7}$   & 2.19954                                      & 0.02\%                                                                  \\* 
			\hline
			\multirow{6}{*}{LS}  & $10^{2}$   & 2.31566                                      & 5.26\%                                                                  \\*
			& $10^{3}$   & 2.22096                                      & 0.95\%                                                                  \\*
			& $10^{4}$   & 2.15422                                      & 2.08\%                                                                  \\*
			& $10^{5}$   & 2.18542                                      & 0.66\%                                                                  \\*
			& $10^{6}$   & 2.20057                                      & 0.03\%                                                                  \\*
			& $10^{7}$   & 2.19946                                      & 0.02\%                                                                  \\* 
			\hline
			\multirow{6}{*}{MoM} & $10^{2}$   & 2.62537                                      & 19.34\%                                                                 \\*
			& $10^{3}$   & 2.28164                                      & 3.71\%                                                                  \\*
			& $10^{4}$   & 2.18505                                      & 0.68\%                                                                  \\*
			& $10^{5}$   & 2.19049                                      & 0.43\%                                                                  \\*
			& $10^{6}$   & 2.19942                                      & 0.03\%                                                                  \\*
			& $10^{7}$   & 2.20004                                      & 0.00\%                                                                  \\* 
			\hline
			\multirow{6}{*}{PM}  & $10^{2}$   & 2.7422                                       & 24.65\%                                                                 \\*
			& $10^{3}$   & 2.27278                                      & 3.31\%                                                                  \\*
			& $10^{4}$   & 2.20158                                      & 0.07\%                                                                  \\*
			& $10^{5}$   & 2.18126                                      & 0.85\%                                                                  \\*
			& $10^{6}$   & 2.19776                                      & 0.10\%                                                                  \\*
			& $10^{7}$   & 2.20044                                      & 0.02\%                                                                  \\
			\hline    
		\end{tabular}
		\caption{\label{tab:alpha2.2} Sample results for $\alpha$ = 2.2 (using sample 1 from each run). The sample sizes are $10^3, 10^4, 10^5, 10^6 $ and $10^7$.}
	\end{table}

	\subsection{Accuracy of Hill's Estimator - Results}
	The HE method differs from the other methods in this paper as one needs to specify a cut-off point from which to consider the ``tail'' portion of the distribution. This is denoted by the \code{k} value which is the \textit{$k^{th}$} largest observation of the data set. In other words, it can be thought of as the number of observations of the tail to be considered in estimating the shape parameter (tail index). We may also specify a value as a starting point of the tail estimation area instead of a ranked order.
	
	\subsection{Hill's Estimator for Strictly Pareto data}
	In this section we look at the impact of \code{k} when HE is used on strictly Pareto data. We consider the same random Pareto samples (with $\alpha = 0.5, 1.5, 2.2$ and $5.0$) used for the previous estimators but only consider $10^3, 10^5$ and $10^6$ sample sizes along with \code{k} set at the $25^{th}$, $50^{th}$ and $75^{th}$ percentiles for each sample size. (Refer Appendix \ref{sec:App-Hill} for the full set of results.)
	
\begin{figure}[h]
		
		\begin{subfigure}{0.5\textwidth}
			\includegraphics[width=0.9\linewidth, height=5cm]{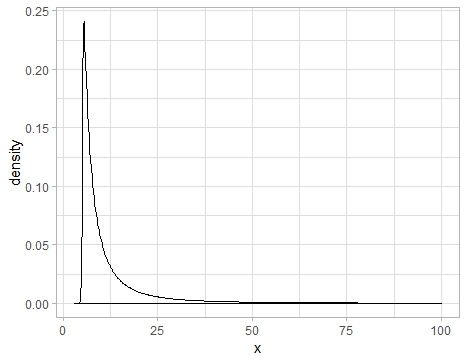} 
			\caption{Density plot of Pareto distribution.}
			\label{fig:pareto-density}
		\end{subfigure}
		\begin{subfigure}{0.5\textwidth}
			\includegraphics[width=0.9\linewidth, height=5cm]{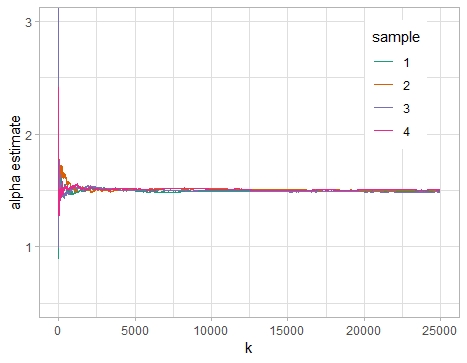}
			\caption{Hill's estimate for $\alpha$ with varying \code{k}.}
			\label{fig:pareto-k}
		\end{subfigure}
		\caption{Plot of pdf of Pareto distribution ($\alpha=1.5$, $\hat{x}_{\mathrm{min}} = 5$) and Hill's estimate for varying \code{k}.}
	\end{figure}

	We also generate four samples from the Pareto distribution with $\alpha=1.5$ and  $\hat{x}_{\mathrm{min}} = 5$. Figure \ref{fig:pareto-k} shows how the HE varies with \code{k}. Since the data supplied is strictly Pareto, increasing \code{k} (from 1 to \code{n} = 100,000) increases the accuracy of the estimate and we can see the estimate stabilizing around $\hat{\alpha} = 1.5$.

	\subsection{Hill's Estimator for Non-Pareto data}
	To further test the accuracy of the HE method and understand the relationship of the value of \code{k} in estimating $\alpha$, we vary \code{k} in the presence of non-Pareto power law data. We consider data sampled from a i) t-distribution and ii) stable distribution respectively, and estimate the tail-index.
	
\begin{figure}[h]
		
		\begin{subfigure}{0.5\textwidth}
			\includegraphics[width=0.9\linewidth, height=5cm]{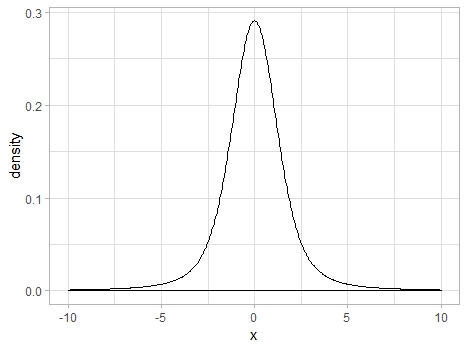} 
			\caption{Density plot of t-distribution.}
			\label{fig:t-density}
		\end{subfigure}
		\begin{subfigure}{0.5\textwidth}
			\includegraphics[width=0.9\linewidth, height=5cm]{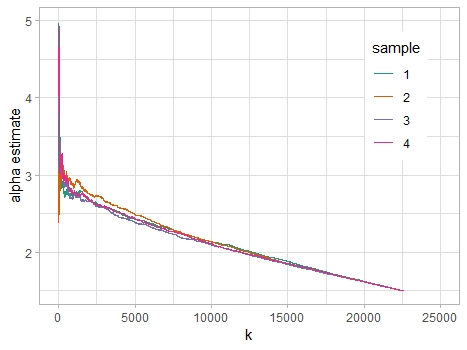}
			\caption{Hill's estimate for $\alpha$ with varying \code{k}.}
			\label{fig:t}
		\end{subfigure}
		\caption{Plot of pdf of t-distribution (d.f. = 3) and Hill's estimate for varying \code{k}.}
	\end{figure}

	For this purpose we first generate four samples from the t-distribution with degrees of freedom (d.f.) equal to 3. The tail index $\alpha$ is estimated by varying \code{k} from 1 to \code{n} (where \code{n} = 100,000). It can be seen from Figure \ref{fig:t} that using a smaller value of \code{k} results in a more accurate estimation of the tail index - we feel this is because a smaller value of \code{k} avoids including observations from the ``body'' of the distribution. We should note however the earlier point about the bias of the estimator for small \code{k}.
	
	Next we simulate four samples from a symmetric stable distribution\footnote{A symmetric stable distribution was generated using $\frac{\sin(\alpha V)}{(\cos(V))^{1/\alpha}}(\frac{\cos((1-\alpha)V)}{W})^{(1-\alpha)/\alpha} $ where $V \sim Unif[-\pi/2,\pi/2]$ and $W \sim Exp(1)$ \citep[p.149]{Glas04}.} with stability parameter\footnote{A symmetric stable distribution with stability parameter $\alpha$ has the following property $\lim_{x \rightarrow \pm \infty} p(x) \sim |x|^{1+\alpha}$.} (tail index) $\alpha$=1.5 and vary \code{k} to estimate $\alpha$. The results of this simulation are similar to that of t-distribution in that a smaller value of \code{k} leads to a closer estimate of the true $\alpha$. The results are as follows:
	
	\begin{figure}[h]
		
		\begin{subfigure}{0.5\textwidth}
			\includegraphics[width=0.9\linewidth, height=5cm]{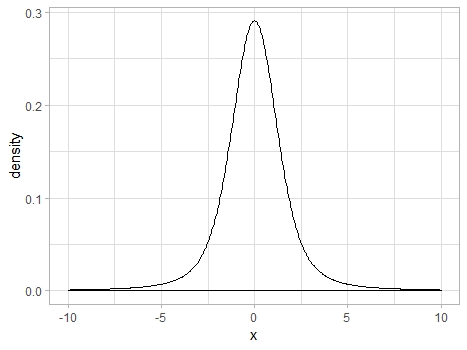} 
			\caption{Density plot of stable distribution.}
			\label{fig:stable-density}
		\end{subfigure}
		\begin{subfigure}{0.5\textwidth}
			\includegraphics[width=0.9\linewidth, height=5cm]{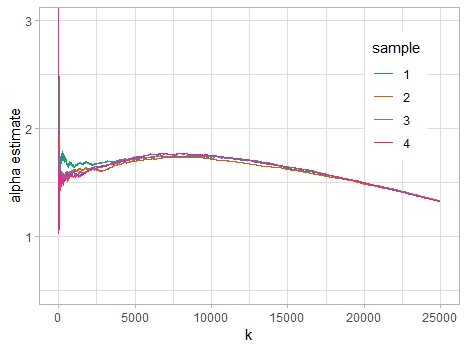}
			\caption{Hill's estimate for $\alpha$ with varying \code{k}.}
			\label{fig:stable}
		\end{subfigure}
		\caption{Plot of pdf of stable distribution ($\alpha = 1.5$) and Hill's estimate for varying \code{k}.}
	\end{figure}
	
	\newpage
	\section{Using the ptsuite Package} \label{sec:usingthepackage}
	The \pkg{ptsuite} is built with ease of use in mind. Here we go over the use of each of the functions provided in the package. The \pkg{ptsuite} is available on Centralized R Archive Network (CRAN): \url{https://cran.r-project.org/web/packages/ptsuite/index.html}.
	
	\subsection[The generate-pareto function]{The \code{generate\_pareto} function}
	This function is able to generate random Pareto distributed data with the specified \code{shape} and \code{scale} parameters. The function has been written to be similar in type to the popular \code{runif} and \code{rexp} type of functions for generating data from a particular distribution.
	
	The full call of the function is:
	\begin{Code}
		generate_pareto(sampleSize, shape, scale)
	\end{Code}
	
	For example to generate a sample of size $100,000$ with $\alpha$ (shape parameter) = 1.2 and $x_{min}$ (scale parameter) = 3 and store it in the variable \code{data}, the following code could be used:
	\begin{CodeChunk}
		\begin{CodeInput}
		R> set.seed(1234)
		R> d <- generate_pareto(100000, 1.2, 3)
		R> head(d)
		\end{CodeInput}
		
		This produces the following output:
		\begin{CodeOutput}
		[1] 18.364397  4.454392  4.533604  4.447960  3.398763  
		4.349730
		\end{CodeOutput}
	\end{CodeChunk}
	
	\subsection[The pareto_qq_test function]{The \code{pareto\_qq\_test} function}
	The \code{pareto\_qq\_test} function can be used as a first step to identify whether the data is Pareto distributed before estimating the tail index. If most of the data points appear to be distributed along a line, it is possible that the data may be Pareto. Conversely, if most of the data are distributed non-linearly, then the data is most probably not Pareto distributed.
	
	This function\footnote{For added interactivity this package makes use of the \pkg{plotly} package \citep{plotlyR} to generate the Q-Q plots if it is available in the user's \proglang{R} library. If unavailable, it defaults to the \proglang{R} base plot.} plots the quantiles of the standard exponential distribution on the x-axis and the log values of the provided data on the y-axis. If Pareto data was supplied, a log transformation of this data would result in an exponential distribution with mean $\alpha^{-1}$. These data points would then show up on the QQ-plot as line with slope $\alpha^{-1}$.
	\begin{leftbar}
		\textbf{Note:} This is a heuristic test where one can look for data being distributed along a straight line indicating a possible Pareto distribution.
	\end{leftbar}
	
	The full call of the function is:
	\begin{Code}
		pareto_qq_test(dat)
	\end{Code}
	
	Demonstration of this function is carried out using a Pareto distributed sample of size $100,000$ with $\alpha$ (shape parameter) = 1.2 and $x_{min}$ (scale parameter) = 3, and using an exponentially distributed sample of size $100,000$ with $\lambda$ (reciprocal of mean) = 5 as an example of non-Pareto data.\\
	
	Q-Q Plot constructed for Pareto distributed data:
	\begin{CodeChunk}
		\begin{CodeInput}
		R> set.seed(1234)
		R> d <- generate_pareto(100000, 1.2, 3)
		R> pareto_qq_test(d)
		\end{CodeInput}
	\end{CodeChunk}
	
	Q-Q Plot constructed for exponentially distributed data:
	\begin{CodeChunk}
		\begin{CodeInput}
		R> set.seed(1234)
		R> exp_data <- rexp(100000, 5)
		R> pareto_qq_test(exp_data)
		\end{CodeInput}
	\end{CodeChunk}
	
	\begin{figure}[h]
		\begin{subfigure}{0.5\textwidth}
			\includegraphics[width=0.9\linewidth, height=5cm]{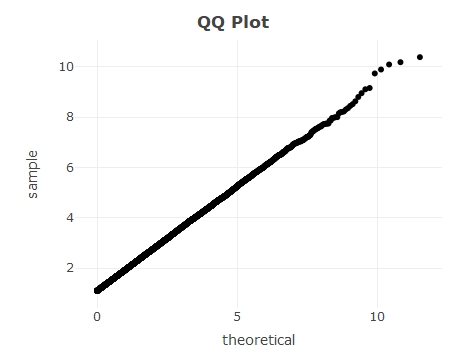} 
			\caption{Q-Q Plot for Pareto distributed data.}
			\label{fig:QQ_pareto}
		\end{subfigure}
		\begin{subfigure}{0.5\textwidth}
			\includegraphics[width=0.9\linewidth, height=5cm]{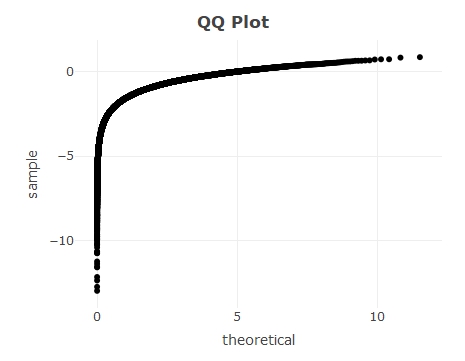}
			\caption{Q-Q Plot for exponentially distributed data.}
			\label{fig:QQ_exp}
		\end{subfigure}
		\caption[Outputs of pareto_qq_test function.]{Outputs of \code{pareto\_qq\_test} function.}
	\end{figure}
	
	\subsection[The pareto_test function]{The \code{pareto\_test} function}
	The \code{pareto\_test} function can be used to identify whether the data is Pareto distributed  \citep{Gulati}. The test generates a p-value corresponding to the actual distribution of the data and is tested for significance. In the case of Pareto data, the p-value should be greater than the pre-determined significance level (generally taken as $0.05$). In addition to using the function on Pareto data, we tested the function on selected non-Pareto (generated) data sets to ensure the test would reject those as Pareto. The tests were conducted on generated data sets of various sizes. The results\footnote{In the table, half-normal distribution is taken to be the positive tail of the standard normal distribution excluding zeros.} obtained are shown in Table \ref{tab:pvalues}.
	
	\begin{table}[!h]
		\centering
		\begin{tabular}{lcr} 
			\hline
			\textbf{Distribution}                                     & \multicolumn{1}{l}{\textbf{Sample Size}} & \multicolumn{1}{l}{\textbf{p-value}}  \\ 
			\hline
			\multirow{3}{*}{Pareto (shape = 1.2, scale = 3)} & $10^{2}$                             & 0.5056804                    \\
			& $10^{3}$                            & 0.1595162                    \\
			& $10^{5}$                          & 0.849023                     \\ 
			\hline
			\multirow{3}{*}{Exponential (rate = 5)}          & $10^{2}$                             & $4.76\times10^{-43}$                     \\
			& $10^{3}$                            & 0                            \\
			& $10^{5}$                          & 0                            \\ 
			\hline
			\multirow{3}{*}{Binomial (n = 20, p = 0.6)}      & $10^{2}$                             & $1.06\times10^{-44}$                     \\
			& $10^{3}$                            & 0                            \\
			& $10^{5}$                          & 0                            \\
			\hline
			\multirow{3}{*}{Normal (mean = 5, std = 3); positive tail} & $10^{2}$                             & $1.26\times10^{-104}$                    \\*
			& $10^{3}$                            & 0                            \\*
			& $10^{5}$                          & 0                            \\
			\hline
		\end{tabular}
		\caption{\label{tab:pvalues} Results of \code{pareto\_test} function for various sample sizes.}
	\end{table}
	
	The full call of the function is:
	\begin{Code}
		pareto_test(dat)
	\end{Code}
	
	Demonstration of this function is carried out using a Pareto distributed sample of size $100,000$ with $\alpha$ (shape parameter) = 1.2 and $x_{min}$ (scale parameter) = 3, and using an exponentially distributed sample of size $100,000$ with $\lambda$ (reciprocal of mean) = 5 as an example of non-Pareto data.\\
	
	\begin{CodeChunk}	
		\begin{CodeInput}
		R> set.seed(1234)
		R> d <- generate_pareto(100000, 1.2, 3)
		R> pareto_test(d)
		\end{CodeInput}
	
		\begin{CodeOutput}
		$`p-value`
		[1] 0.8604162	
		\end{CodeOutput}
		
		\begin{CodeInput}
		R> set.seed(1234)
		R> exp_data <- rexp(100000, 5)
		R> pareto_test(exp_data)
		\end{CodeInput}
		\begin{CodeOutput}
		$`p-value`
		[1] 0
		\end{CodeOutput}
	\end{CodeChunk}

	\subsection[The alpha_mle function]{The \code{alpha\_mle} function}
	This function can be used to estimate the shape parameter ($\alpha$) using the Maximum Likelihood Estimator method - please refer equation (\ref{eq:MLE}). It can be used to obtain biased and unbiased estimates of the shape and scale parameters as well as the confidence interval for the shape parameter for the biased estimates.
	
	The full call of the function is:
	\begin{Code}
		alpha_mle(dat, biased = TRUE, significance = NULL)
	\end{Code}
	
	To demonstrate the various possibilities of this function we first generate Pareto data with $\alpha = 1.2$ and $x_{min}=3 $ and assign it to the variable \code{data}. \code{alpha\_mle} is then used to obtain:
	\begin{enumerate}
		\item Biased Estimates for $\alpha$ and $x_{min}$
		\item Unbiased Estimates for $\alpha$ and $x_{min}$
		\item Biased Estimates for $\alpha$ and $x_{min}$ and the $95\%$ confidence interval for $\alpha$
	\end{enumerate}
	
	R implementation of the above mentioned demonstration is given below.
	\\\\Generating data from a Pareto Distribution:
	\begin{CodeChunk}	
		\begin{CodeInput}
		R> set.seed(1234)
		R> d <- generate_pareto(100000, 1.2, 3)
		\end{CodeInput}
		Obtaining the biased maximum likelihood estimates for $\alpha$ and $x_{min}$:
		\begin{CodeInput}
		R> alpha_mle(d)
		\end{CodeInput}
		\begin{CodeOutput}
		$shape
		[1] 1.201384
			
		$scale
		[1] 3.000005
			
		\end{CodeOutput}
		Obtaining the unbiased maximum likelihood estimates for $\alpha$ and $x_{min}$:
		\begin{CodeInput}
		R> alpha_mle(d, FALSE)
		\end{CodeInput}
		\begin{CodeOutput}
		$shape
		[1] 1.201359
			
		$scale
		[1] 2.99998
			
		\end{CodeOutput}
		Obtaining the biased maximum likelihood estimator for $\alpha$ and $x_{min}$ and the corresponding $95\%$ confidence interval for $\alpha$:
		\begin{CodeInput}
		R> alpha_mle(d, TRUE, 0.05)
		\end{CodeInput}
		\begin{CodeOutput}
		$shape
		[1] 1.201384
			
		$lower_bound
		[1] 1.193937
			
		$upper_bound
		[1] 1.20883
			
		$scale
		[1] 3.000005
			
			
		\end{CodeOutput}
	\end{CodeChunk}
	
	\subsection[The alpha_hills function]{The \code{alpha\_hills} function}
	The Hill's Estimator - please refer equation (\ref{eq:Hill}) - is particularly useful in the fact that it allows the specification of \code{k} as either the number of observations to be included in the tail (where \code{value = FALSE}) or the minimum value to be considered in the tail of the distribution (where \code{value = TRUE}). When \code{k}=\code{n}, the Hill's Estimator returns the same estimate as \code{alpha_mle} with a warning notifying the user.
	
	The full call of the function is:
	\begin{Code}
		alpha_hills(dat, k, value = FALSE)
	\end{Code}
	
	We demonstrate the usage of the Hill's Estimator by generating Pareto data with $\alpha$=1.2 and $x_{min}$=3, assigning it to the variable \code{data} and then using \code{alpha_hills}
	
	R implementation of the above mentioned demonstration is given below with the respective outputs:
	\begin{CodeChunk}
		\begin{CodeInput}
		R> set.seed(1234)
		R> d <- generate_pareto(100000, 1.2, 3)
		R> alpha_hills(d, 8000, FALSE)
		\end{CodeInput}
		\begin{CodeOutput}
		$shape
		[1] 1.198717
			
		$scale
		[1] 24.54499
			
			
		\end{CodeOutput}
		\begin{CodeInput}		
		R> alpha_hills(d, 5000, TRUE)
		\end{CodeInput}
		\begin{CodeOutput}
		$shape
		[1] 1.125696
			
		$scale
		[1] 5002.76
			
		\end{CodeOutput}
		\begin{CodeInput}		
		R> alpha_hills(d, 100000, FALSE)
		\end{CodeInput}
		\begin{CodeOutput}
		$shape
		[1] 1.201384
			
		$scale
		[1] 3.000005
			
		Warning message:
		In alpha_hills(d, 1e+05, FALSE) :
		Setting k as the number of observations makes it equivalent 
		to the MLE (alpha_mle function).
		\end{CodeOutput}
	\end{CodeChunk}

	\subsection[The alpha_ls function]{The \code{alpha\_ls} function}
	This function can be used to estimate the shape parameter ($\alpha$) using the Least Squares Estimator method - please refer equation (\ref{eq:LS1}).
	
	The full call of the function is:
	\begin{Code}
		alpha_ls(dat)
	\end{Code}
	
	We demonstrate the usage of the Least Squares Estimator by generating Pareto data with $\alpha$=1.2 and $x_{min}$=3, assigning it to the variable \code{data} and then using \code{alpha_ls}.
	
	R implementation of the above mentioned demonstration is given below with the respective output:
	\begin{CodeChunk}
		\begin{CodeInput}
		R> set.seed(1234)
		R> d <- generate_pareto(100000, 1.2, 3)
		R> alpha_ls(d)
		\end{CodeInput}
		
		\begin{CodeOutput}
		$shape
		[1] 1.20067
			
		$scale
		[1] 3.000005
			
		\end{CodeOutput}
		
	\end{CodeChunk}
	
	\subsection[The alpha_percentile function]{The \code{alpha\_percentile} function}
	This function can be used to estimate the shape parameter ($\alpha$) using the Percentile Estimator method - please refer equation (\ref{eq:PM}).
	
	The full call of the function is:
	\begin{Code}
		alpha_percentile(dat)
	\end{Code}
	
	We demonstrate the usage of the Percentile Estimator by generating Pareto data with $\alpha$=1.2 and $x_{min}$=3, assigning it to the variable \code{data} and then using \code{alpha_percentile}.
	
	R implementation of the above mentioned demonstration is given below with the respective output:
	\begin{CodeChunk}
		\begin{CodeInput}
		R> set.seed(1234)
		R> d <- generate_pareto(100000, 1.2, 3)
		R> alpha_percentile(d)
		\end{CodeInput}
		
		\begin{CodeOutput}
		$shape
		[1] 1.20048
			
		$scale
		[1] 3.000005	
			
		\end{CodeOutput}
		
	\end{CodeChunk}

	\subsection[The alpha_modified_percentile function]{The \code{alpha\_modified\_percentile} function}
	This function can be used to estimate the shape parameter ($\alpha$) using the Modified Percentile Estimator method - please refer equation (\ref{eq:MPM}).
	
	The full call of the function is:
	\begin{Code}
		alpha_modified_percentile(dat)
	\end{Code}
	
	We demonstrate the usage of the Modified Percentile Estimator by generating Pareto data with $\alpha$=1.2 and $x_{min}$=3, assigning it to the variable \code{data} and then using\\ \code{alpha_modified_percentile}.
	
	R implementation of the above mentioned demonstration is given below with the respective output:
	\begin{CodeChunk}
		\begin{CodeInput}
		R> set.seed(1234)
		R> d <- generate_pareto(100000, 1.2, 3)
		R> alpha_modified_percentile(d)
		\end{CodeInput}
		
		\begin{CodeOutput}
		$shape
		[1] 1.196936
			
		$scale
		[1] 3.000005
			
		\end{CodeOutput}
		
	\end{CodeChunk}
	
	\subsection[The alpha_geometric_percentile function]{The \code{alpha\_geometric\_percentile} function}
	This function can be used to estimate the shape parameter ($\alpha$) using the Geometric Percentile Estimator method - please refer equation (\ref{eq:GMPM}).
	
	The full call of the function is:
	\begin{Code}
		alpha_geometric_percentile(dat)
	\end{Code}
	
	We demonstrate the usage of the Geometric Percentile Estimator by generating Pareto data with $\alpha$=1.2 and $x_{min}$=3, assigning it to the variable \code{data} and then using\\ \code{alpha_geometric_percentile}.
	
	R implementation of the above mentioned demonstration is given below with the respective output:
	\begin{CodeChunk}
		\begin{CodeInput}
		R> set.seed(1234)
		R> d <- generate_pareto(100000, 1.2, 3)
		R> alpha_geometric_percentile(d)
		\end{CodeInput}
		
		\begin{CodeOutput}
		$shape
		[1] 1.195801
			
		$scale
		[1] 3.000005
			
			
		\end{CodeOutput}
		
	\end{CodeChunk}
	
	\subsection[The alpha_wls function]{The \code{alpha\_wls} function}
	This function can be used to estimate the shape parameter ($\alpha$) using the Weighted Least Squares Estimator method - please refer to equations (\ref{eq:WLS1}) and (\ref{eq:WLS2}).
	
	The full call of the function is:
	\begin{Code}
		alpha_wls(dat)
	\end{Code}
	
	We demonstrate the usage of the Weighted Least Squares Estimator by generating Pareto data with $\alpha$=1.2 and $x_{min}$=3, assigning it to the variable \code{data} and then using \code{alpha_wls}.
	
	R implementation of the above mentioned demonstration is given below with the respective output:
	\begin{CodeChunk}
		\begin{CodeInput}
		R> set.seed(1234)
		R> d <- generate_pareto(100000, 1.2, 3)
		R> alpha_wls(d)
		\end{CodeInput}
		\begin{CodeOutput}
		$shape
		[1] 1.201303
			
		$scale
		[1] 3.000005
			
		\end{CodeOutput}
		
	\end{CodeChunk}
	
	\subsection[The alpha_moment function]{The \code{alpha\_moment} function}
	This function can be used to estimate the shape parameter ($\alpha$) using the Moment Estimator method - please refer equation (\ref{eq:MoM}).
	
	The full call of the function is:
	\begin{Code}
		alpha_moment(dat)
	\end{Code}
	
	We demonstrate the usage of the Moment Estimator by generating Pareto data with $\alpha$=1.2 and $x_{min}$=3, assigning it to the variable \code{data} and then using \code{alpha_moment}.
	
	R implementation of the above mentioned demonstration is given below with the respective output:
	\begin{CodeChunk}
		\begin{CodeInput}
		R> set.seed(1234)
		R> d <- generate_pareto(100000, 1.2, 3)
		R> alpha_moment(d)
		\end{CodeInput}
		\begin{CodeOutput}
		$shape
		[1] 1.241772
			
		$scale
		[1] 3.000005		
			
		\end{CodeOutput}
		
	\end{CodeChunk}
	
	\subsection[The generate_all_estimates function]{The \code{generate\_all\_estimates} function}
	This function can be used to obtain estimates of the shape parameter ($\alpha$) using all estimators except the HE.
	
	The full call of the function is:
	\begin{Code}
		generate_all_estimates(dat)
	\end{Code}
	
	We demonstrate the usage of the function by generating Pareto data with $\alpha$=1.2 and $x_{min}$=3, assigning it to the variable \code{data} and then using \code{generate_all_estimates}.
	
	R implementation of the above mentioned demonstration is given below with the respective output:
	\begin{CodeChunk}
		\begin{CodeInput}
	R> set.seed(1234)
	R> d <- generate_pareto(100000, 1.2, 3)
	R> generate_all_estimates(d)
		\end{CodeInput}
		\begin{CodeOutput}
	Method.of.Estimation Shape.Parameter Scale.Parameter
	1  Maximum Likelihood Estimate        1.201384        3.000005
	2                Least Squares        1.200670        3.000005
	3            Method of Moments        1.241772        3.000005
	4           Percentiles Method        1.200480        3.000005
	5  Modified Percentiles Method        1.196936        3.000005
	6 Geometric Percentiles Method        1.195801        3.000005
	7       Weighted Least Squares        1.201303        3.000005		
			
		\end{CodeOutput}
		
	\end{CodeChunk}
	
	
	\section{Summary and Future Work} \label{sec:summary}
	
	\begin{leftbar}
		Our \proglang{R} package \pkg{ptsuite} can be used in tail index estimation for Pareto distributed data. One may use the \code{pareto-qq-test} function as a first step to test (heuristically) if the data is Pareto. There are a number of methods which can be called on to estimate the tail index, and it is advisable to use more than one method as a comparison for the estimate.
		
		Our main focus in developing this package was on 
		run-times for parameter estimation for large data samples. The estimation methods on \pkg{ptsuite} were faster than their counterpart methods in the packages \pkg{laeken} and \pkg{EnvStats}. The Pareto data generation function on \pkg{ptsuite} was also faster than that of the package \pkg{EnvStats}. All functions on \pkg{ptsuite} performed reasonably well in the speed tests, where we tested data sets with up to $10^7$ data points. 
		
		For future work we hope to add more tail estimation methods and functions found in the references to this package. In particular we hope to eventually incorporate the vast catalog of estimators in reference \cite{Hundred}.
	\end{leftbar}

	
	\section*{Computational details}
	
	The results in this paper were obtained using 
	\proglang{R}~3.5.2. 
	and Microsoft \proglang{R}~3.5.1 \citep{micr}\footnote{Microsoft \proglang{R} 
		was used for the benchmark results of the speed and we note here that this version of \proglang{R} appeared to use slightly lower times in running the same tests.} with the
	\pkg{ptsuite}~1.0.0 package. \proglang{R} itself
	and all packages used are available from the Comprehensive
	\proglang{R} Archive Network (CRAN) at
	\url{https://CRAN.R-project.org/}. Microsoft \proglang{R} may be obtained from \url{https://mran.microsoft.com/open}.
	
	RStudio V 1.1.423 \citep{rStudio} was used as the IDE for all tasks related to package development and generating results. The packages \pkg{devtools} \citep{devtools} and \pkg{roxygen2} \citep{roxygen} were used to assist in package documentation and development. The \pkg{Rcpp} package \citep{rcpp1,rcpp2,rcpp3} was used in the compilation of \proglang{C++} code used in all estimation functions of the package.
	
	\pkg{microbenchmark} 1.4-4, \pkg{ggplot2} 3.0.0 \citep{ggplot}  and \pkg{plotly} 4.8.0 were used in generating the results included in this paper.
	
	Two computers were used in the building of this package and the generation of the results. The technical specifications of the two computers are as follows:
	
	\begin{table}[h]
		\centering
		\begin{tabular}{lll} 
			\hline
			\textbf{Specification} & \textbf{Laptop}                                                                                                                & \textbf{Desktop}                                                                                                               \\ 
			\hline
			\textbf{CPU}           & \begin{tabular}[c]{@{}l@{}}Intel(R) Core(TM) i5-8250U~\\@ 1.60GHz, 1800 Mhz, \\4 Core(s), 8 Logical Processor(s) \end{tabular} & \begin{tabular}[c]{@{}l@{}}Intel(R) Core(TM) i7-7700~\\@ 3.60GHz, 3601 Mhz, \\4 Core(s), 8 Logical Processor(s) \end{tabular}  \\
			\textbf{RAM}           & 8 GB                                                                                                                           & 32 GB                                                                                                                          \\
			\textbf{OS}            & \begin{tabular}[c]{@{}l@{}}Windows 10 Enterprise \\LTSC 10.0.17763 Build 17763\end{tabular}                                    & \begin{tabular}[c]{@{}l@{}}Windows 10 Pro \\10.0.17763 Build 17763\end{tabular} \\\hline                                            
		\end{tabular}
	\end{table}

	\section*{Acknowledgments}
	
	\begin{leftbar}
		We wish to acknowledge Dr. J. Nair for helpful advice and in providing unpublished material for our research and Prof. S. Gulati for sharing knowledge on the Goodness of Fit tests. D. J. would also like to acknowledge Prof. S. Banneheka for a very useful discussion on the WLS method of estimation.
	\end{leftbar}

	
	\bibliography{references_PL}

	
	\newpage
	\begin{appendix}
		
		\section{Complete Tables of Estimates} \label{sec:App-Est}
		
		\begin{footnotesize}
			
\begin{longtable}{lccrrr} 
				\multicolumn{6}{r}{\tablename\ \thetable\ -- \textit{Continued on next page}} \\
				\endfoot
				\endlastfoot
				\hline
				\textbf{Method}             & \textbf{Sample Size}                  & \multicolumn{1}{l}{\textbf{Sample}} & \multicolumn{1}{c}{$\hat{\alpha}$ } & \multicolumn{1}{c}{$\hat{x}_{\mathrm{min}}$ } & \multicolumn{1}{c}{$\hat{\alpha}$ \textbf{error} }  \\* 
				\hline
				\endhead
				\multirow{12}{*}{MLE}  & \multirow{4}{*}{$10^{3}$}    & 1                           & 5.0652128                            & 2.000028                                       & 1.30\%                                      \\*
				&                          & 2                           & 4.8768226                            & 2.000292                                       & 2.46\%                                      \\*
				&                          & 3                           & 5.1345472                            & 2.000087                                       & 2.69\%                                      \\*
				&                          & 4                           & 4.7497352                            & 2.000301                                       & 5.01\%                                      \\* 
				\cline{2-6}
				& \multirow{4}{*}{$10^{5}$}  & 1                           & 4.9810971                            & 2.000021                                       & 0.38\%                                      \\*
				&                          & 2                           & 5.0083374                            & 2.000011                                       & 0.17\%                                      \\*
				&                          & 3                           & 4.9807569                            & 2.000002                                       & 0.38\%                                      \\*
				&                          & 4                           & 5.0034707                            & 2.000001                                       & 0.07\%                                      \\
\end{longtable}
			
			\newpage
			\begin{longtable}{lccrrr} 
				\multicolumn{6}{r}{\tablename\ \thetable\ -- \textit{Continued on next page}} \\
				\endfoot
				\endlastfoot
				\hline
				\textbf{Method}             & \textbf{Sample Size}                  & \multicolumn{1}{l}{\textbf{Sample}} & \multicolumn{1}{c}{$\hat{\alpha}$ } & \multicolumn{1}{c}{$\hat{x}_{\mathrm{min}}$ } & \multicolumn{1}{c}{$\hat{\alpha}$ \textbf{error} }  \\* 
				\hline
				\endhead
				\multirow{12}{*}{MLE}  & \multirow{4}{*}{$10^{3}$}    & 1                           & 2.2286936                            & 2.000063                                       & 1.30\%                                      \\*
				&                          & 2                           & 2.1458019                            & 2.000664                                       & 2.46\%                                      \\*
				&                          & 3                           & 2.2592007                            & 2.000197                                       & 2.69\%                                      \\*
				&                          & 4                           & 2.0898835                            & 2.000683                                       & 5.01\%                                      \\* 
				\cline{2-6}
				& \multirow{4}{*}{$10^{5}$}  & 1                           & 2.1916827                            & 2.000048                                       & 0.38\%                                      \\*
				&                          & 2                           & 2.2036685                            & 2.000025                                       & 0.17\%                                      \\*
				&                          & 3                           & 2.1915330                            & 2.000005                                       & 0.38\%                                      \\*
				&                          & 4                           & 2.2015271                            & 2.000002                                       & 0.07\%                                      \\
\end{longtable}
			
			\newpage
			\begin{longtable}{lccrrr} 
				\multicolumn{6}{r}{\tablename\ \thetable\ -- \textit{Continued on next page}} \\
				\endfoot
				\endlastfoot
				\hline
				\textbf{Method}             & \textbf{Sample Size}                  & \multicolumn{1}{l}{\textbf{Sample}} & \multicolumn{1}{c}{$\hat{\alpha}$ } & \multicolumn{1}{c}{$\hat{x}_{\mathrm{min}}$ } & \multicolumn{1}{c}{$\hat{\alpha}$ \textbf{error} }  \\* 
				\hline
				\endhead
				\multirow{12}{*}{MLE}  & \multirow{4}{*}{$10^{3}$}    & 1                           & 1.5195638                            & 2.000093                                       & 1.30\%                                      \\*
				&                          & 2                           & 1.4630468                            & 2.000974                                       & 2.46\%                                      \\*
				&                          & 3                           & 1.5403641                            & 2.000289                                       & 2.69\%                                      \\*
				&                          & 4                           & 1.4249206                            & 2.001002                                       & 5.01\%                                      \\* 
				\cline{2-6}
				& \multirow{4}{*}{$10^{5}$}  & 1                           & 1.4943291                            & 2.000071                                       & 0.38\%                                      \\*
				&                          & 2                           & 1.5025012                            & 2.000037                                       & 0.17\%                                      \\*
				&                          & 3                           & 1.4942271                            & 2.000007                                       & 0.38\%                                      \\*
				&                          & 4                           & 1.5010412                            & 2.000002                                       & 0.07\%                                      \\
\end{longtable}
			
			\newpage
			\begin{longtable}{lccrrr} 
				\multicolumn{6}{r}{\tablename\ \thetable\ -- \textit{Continued on next page}} \\
				\endfoot
				\endlastfoot
				\hline
				\textbf{Method}             & \textbf{Sample Size}                  & \multicolumn{1}{l}{\textbf{Sample}} & \multicolumn{1}{c}{$\hat{\alpha}$ } & \multicolumn{1}{c}{$\hat{x}_{\mathrm{min}}$ } & \multicolumn{1}{c}{$\hat{\alpha}$ \textbf{error} }  \\* 
				\hline
				\endhead
				\multirow{12}{*}{MLE}  & \multirow{4}{*}{$10^{3}$}    & 1                           & 0.5065213                            & 2.000278                                       & 1.30\%                                      \\*
				&                          & 2                           & 0.4876823                            & 2.002923                                       & 2.46\%                                      \\*
				&                          & 3                           & 0.5134547                            & 2.000866                                       & 2.69\%                                      \\*
				&                          & 4                           & 0.4749735                            & 2.003009                                       & 5.01\%                                      \\* 
				\cline{2-6}
				& \multirow{4}{*}{$10^{5}$}  & 1                           & 0.4981097                            & 2.000212                                       & 0.38\%                                      \\*
				&                          & 2                           & 0.5008337                            & 2.000112                                       & 0.17\%                                      \\*
				&                          & 3                           & 0.4980757                            & 2.000021                                       & 0.38\%                                      \\*
				&                          & 4                           & 0.5003471                            & 2.000007                                       & 0.07\%                                      \\
\end{longtable}
			
		\end{footnotesize}
		
		\newpage
		
		
		\section{Hill's Estimates} \label{sec:App-Hill}
		\begin{footnotesize}
			\begin{longtable}{crcrrr} 
				\multicolumn{6}{r}{\tablename\ \thetable\ -- \textit{Continued on next page}} \\
				\endfoot
				\endlastfoot
				\hline
				\textbf{Sample Size}             & \code{\textbf{k}}                  & \multicolumn{1}{l}{\textbf{Sample}} & \multicolumn{1}{c}{$\hat{\alpha}$ } & \multicolumn{1}{c}{$\hat{x}_{\mathrm{min}}$ } & \multicolumn{1}{c}{$\hat{\alpha}$ \textbf{error} }  \\* 
				\hline
				\endhead
				\multirow{12}{*}{$10^{3}$}    & \multirow{4}{*}{250}    & 1                                  & 5.1384879                           & 2.622409                         & 2.77\%                            \\*
				&                         & 2                                  & 5.0187719                           & 2.688776                         & 0.38\%                            \\*
				&                         & 3                                  & 5.1645221                           & 2.619874                         & 3.29\%                            \\*
				&                         & 4                                  & 4.7188721                           & 2.662987                         & 5.62\%                            \\* 
				\cline{2-6}
				& \multirow{4}{*}{500}    & 1                                  & 5.358747                            & 2.314142                         & 7.17\%                            \\*
				&                         & 2                                  & 4.7658326                           & 2.301751                         & 4.68\%                            \\*
				&                         & 3                                  & 4.9729643                           & 2.276799                         & 0.54\%                            \\*
				&                         & 4                                  & 4.84922                             & 2.32247                          & 3.02\%                            \\
\end{longtable}	
			\begin{longtable}{crcrrr} 
				\hline
				\textbf{Sample Size}             & \code{\textbf{k}}                  & \multicolumn{1}{l}{\textbf{Sample}} & \multicolumn{1}{c}{$\hat{\alpha}$ } & \multicolumn{1}{c}{$\hat{x}_{\mathrm{min}}$ } & \multicolumn{1}{c}{$\hat{\alpha}$ \textbf{error} }  \\* 
				\hline
				\endhead
				\multirow{12}{*}{$10^{3}$}    & \multirow{4}{*}{250}    & 1                                  & 2.2609347                           & 3.702224                         & 2.77\%                            \\*
				&                         & 2                                  & 2.2082596                           & 3.918605                         & 0.38\%                            \\*
				&                         & 3                                  & 2.2723897                           & 3.694097                         & 3.29\%                            \\*
				&                         & 4                                  & 2.0763037                           & 3.833704                         & 5.62\%                            \\* 
				\cline{2-6}
				& \multirow{4}{*}{500}    & 1                                  & 2.3578487                           & 2.786314                         & 7.17\%                            \\*
				&                         & 2                                  & 2.0969663                           & 2.752523                         & 4.68\%                            \\*
				&                         & 3                                  & 2.1881043                           & 2.685175                         & 0.54\%                            \\*
				&                         & 4                                  & 2.1336568                           & 2.809155                         & 3.02\%                            \\
\end{longtable}
			\begin{longtable}{crcrrr} 
				\multicolumn{6}{r}{\tablename\ \thetable\ -- \textit{Continued on next page}} \\
				\endfoot
				\endlastfoot
				\hline
				\textbf{Sample Size}             & \code{\textbf{k}}                  & \multicolumn{1}{l}{\textbf{Sample}} & \multicolumn{1}{c}{$\hat{\alpha}$ } & \multicolumn{1}{c}{$\hat{x}_{\mathrm{min}}$ } & \multicolumn{1}{c}{$\hat{\alpha}$ \textbf{error} }  \\* 
				\hline
				\endhead
				\multirow{12}{*}{$10^{3}$}    & \multirow{4}{*}{250}    & 1                                  & 1.5415464                           & 4.934745                         & 2.77\%                            \\*
				&                         & 2                                  & 1.5056316                           & 5.363466                         & 0.38\%                            \\*
				&                         & 3                                  & 1.5493566                           & 4.918864                         & 3.29\%                            \\*
				&                         & 4                                  & 1.4156616                           & 5.193897                         & 5.62\%                            \\* 
				\cline{2-6}
				& \multirow{4}{*}{500}    & 1                                  & 1.6076241                           & 3.252596                         & 7.17\%                            \\*
				&                         & 2                                  & 1.4297498                           & 3.194905                         & 4.68\%                            \\*
				&                         & 3                                  & 1.4918893                           & 3.08091                          & 0.54\%                            \\*
				&                         & 4                                  & 1.454766                            & 3.291777                         & 3.02\%                            \\
\end{longtable}
			\begin{longtable}{crcrrr} 
				\multicolumn{6}{r}{\tablename\ \thetable\ -- \textit{Continued on next page}} \\
				\endfoot
				\endlastfoot
				\hline
				\textbf{Sample Size}             & \code{\textbf{k}}                  & \multicolumn{1}{l}{\textbf{Sample}} & \multicolumn{1}{c}{$\hat{\alpha}$ } & \multicolumn{1}{c}{$\hat{x}_{\mathrm{min}}$ } & \multicolumn{1}{c}{$\hat{\alpha}$ \textbf{error} }  \\* 
				\hline
				\endhead
				\multirow{12}{*}{$10^{3}$}    & \multirow{4}{*}{250}    & 1                                  & 0.5138488                           & 30.042363                        & 2.77\%                            \\*
				&                         & 2                                  & 0.5018772                           & 38.572395                        & 0.38\%                            \\*
				&                         & 3                                  & 0.5164522                           & 29.753256                        & 3.29\%                            \\*
				&                         & 4                                  & 0.4718872                           & 35.028376                        & 5.62\%                            \\* 
				\cline{2-6}
				& \multirow{4}{*}{500}    & 1                                  & 0.5358747                           & 8.602613                         & 7.17\%                            \\*
				&                         & 2                                  & 0.4765833                           & 8.152935                         & 4.68\%                            \\*
				&                         & 3                                  & 0.4972964                           & 7.311007                         & 0.54\%                            \\*
				&                         & 4                                  & 0.484922                            & 8.917257                         & 3.02\%                            \\
\end{longtable}
			
			\newpage
			
			\section{LS Estimates} \label{sec:App-LS}
			
			\begin{figure}[H]
				\centering 
				\includegraphics{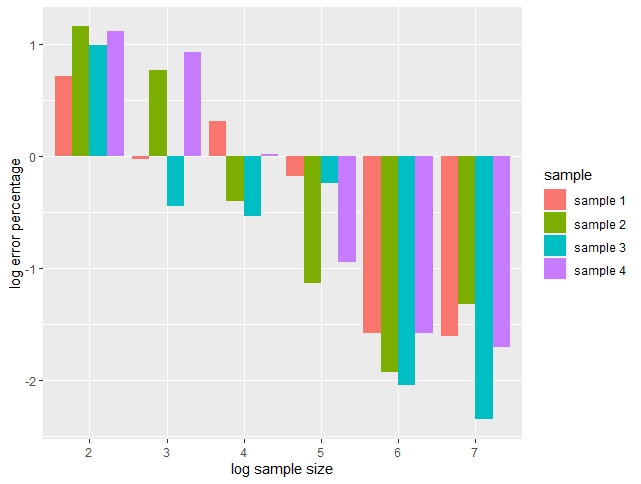}
				\caption{Accuracy plot of LS estimates for distinct samples.}
				\label{fig:LS}
			\end{figure}
			
			\begin{longtable}{ccrr} 
				\multicolumn{4}{r}{\tablename\ \thetable\ -- \textit{Continued on next page}} \\
				\endfoot
				\endlastfoot
				\hline
				\textbf{Sample Size}               & \multicolumn{1}{l}{\textbf{Sample}} & \multicolumn{1}{l}{$\hat{\alpha}$} & \multicolumn{1}{l}{$\hat{\alpha}$ \textbf{error}}  \\* 
				\hline
				\multirow{4}{*}{$10^{2}$} & 1      & 2.31566                                      & 5.26\%                                                                  \\*
				& 2      & 1.88102                                      & 14.50\%                                                                 \\*
				& 3      & 1.98422                                      & 9.81\%                                                                  \\*
				& 4      & 1.91018                                      & 13.17\%                                                                 \\* 
				\hline
				\multirow{4}{*}{$10^{3}$} & 1      & 2.22096                                      & 0.95\%                                                                  \\*
				& 2      & 2.07064                                      & 5.88\%                                                                  \\*
				& 3      & 2.20794                                      & 0.36\%                                                                  \\
\end{longtable}

			\newpage
			
			\section{PM Estimates} \label{sec:App-PM}
			\begin{figure}[H]
				\centering 
				\includegraphics{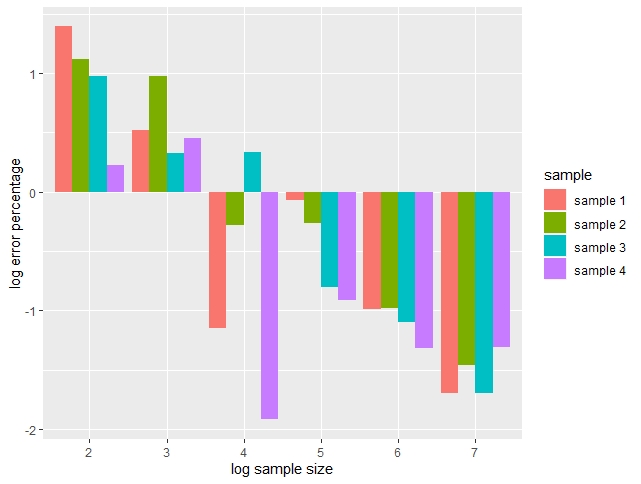} 
				\caption{Accuracy plot of PM estimates for distinct samples.}
				\label{fig:PM}
			\end{figure}
			
			\begin{longtable}{ccrr} 
				\multicolumn{4}{r}{\tablename\ \thetable\ -- \textit{Continued on next page}} \\
				\endfoot
				\endlastfoot
				\hline
				\textbf{Sample Size}               & \multicolumn{1}{l}{\textbf{Sample}} & \multicolumn{1}{l}{$\hat{\alpha}$} & \multicolumn{1}{l}{$\hat{\alpha}$ \textbf{error}}  \\* 
				\hline
				\multirow{4}{*}{$10^{2}$} & 1      & 2.7422                                       & 24.65\%                                                                 \\*
				& 2      & 1.90896                                      & 13.23\%                                                                 \\*
				& 3      & 1.99232                                      & 9.44\%                                                                  \\*
				& 4      & 2.23726                                      & 1.69\%                                                                  \\* 
				\hline
				\multirow{4}{*}{$10^{3}$} & 1      & 2.27278                                      & 3.31\%                                                                  \\*
				& 2      & 1.99423                                      & 9.35\%                                                                  \\*
				& 3      & 2.2465                                       & 2.11\%                                                                  \\
\end{longtable}

			\newpage
			\section{A note on the Generalized Pareto Distribution and evir package} \label{sec:GPDevir}
			
			In this section we discuss briefly the \proglang{R}-package \pkg{evir}. The other packages and their respective functions considered in this paper are straight forward in that they work directly with Pareto distributions for data generation/tail index estimation. The \proglang{R}-package \pkg{evir}, on the other hand, works with the Generalized Pareto Distribution (GPD). The GPD has the following form for its probability density function:
			
			\be \label{eq:GPDpdf}
			g(x) = \frac{1}{\si} \left[ 1 + \frac{\xi (x-\mu)}{\si} \right]^{-(1 + 1/ \xi)}
			\ee 
			The pdf (\ref{eq:GPDpdf}) is defined for $x \geq \mu$ when $\xi > 0$ and is then equivalent to the Pareto distribution when making the following identifications:
			\bea
			x_{\mathrm{min}} &=& \frac{\si}{\xi}  = \mu \\
			\alpha &=& \frac{1}{\xi} \label{eq:tail_gpd} 
			\eea
			To compare our package \pkg{ptsuite} with \pkg{evir} we work with the transformations above. In particular to generate Pareto data, we call the function \code{rgpd} in \pkg{evir} and input $\xi, \si, \mu$ in terms of the corresponding $x_{\mathrm{min}}$ and $\alpha$.  For the tail index evaluation in \pkg{evir}, we call the function \code{gpd}. One of the outputs of \code{gpd} is an estimate of $\xi$ which would need to be inverted to give the tail index as per (\ref{eq:tail_gpd}) . In addition, there is an input  \code{nextremes} for the function  \code{gpd}  whereby the number of data points is specified. The function \code{gpd} uses MLE when \code{nextremes} is set to the sample size. To use HE method we specify \code{nextremes} accordingly, e.g. for a sample size = 1000, setting \code{nextremes} = 990 would omit the ten smallest observations.
			
\begin{longtable}{cccccc} 
	\hline
	\multirow{2}{*}{Sample Size} & \multicolumn{1}{l}{\multirow{2}{*}{$\alpha$}} & \multicolumn{2}{c}{\pkg{ptsuite}}                                                                                                               & \multicolumn{2}{c}{\pkg{EnvStats}}                                                                                                               \\* 
	\cline{3-6}
	& \multicolumn{1}{l}{}                                           & \multicolumn{1}{c}{{$\hat{\alpha}$}} & \multicolumn{1}{c}{{$\hat{\alpha}$ \textbf{error}}} & \multicolumn{1}{c}{{$\hat{\alpha}$}} & \multicolumn{1}{c}{{$\hat{\alpha}$ \textbf{error}}}  \endfirsthead 
	\hline
	\multirow{3}{*}{$10^{3}$}                & 0.5                                                            & 0.50476                                                          & 0.95\%                                                                  & 0.51848                                                          & 3.70\%                                                                   \\
	& 1.5                                                            & 1.51429                                                          & 0.95\%                                                                  & 1.55544                                                          & 3.70\%                                                                   \\
	& 2.2                                                            & 2.22096                                                          & 0.95\%                                                                  & 2.28131                                                          & 3.70\%                                                                   \\ 
	\hline
	\multirow{3}{*}{$10^{5}$}                & 0.5                                                            & 0.49669                                                          & 0.66\%                                                                  & 0.49709                                                          & 0.58\%                                                                   \\
	& 1.5                                                            & 1.49006                                                          & 0.66\%                                                                  & 1.49127                                                          & 0.58\%                                                                   \\
\end{longtable}

			\normalsize
			
			\newpage
			\small
			\begin{longtable}
{cccccc} 
	\hline
\multirow{2}{*}{Sample Size} & \multicolumn{1}{l}{\multirow{2}{*}{$\alpha$}} & \multicolumn{2}{c}{\pkg{ptsuite}}                                                                                                               & \multicolumn{2}{c}{\pkg{EnvStats}}                                                                                                               \\* 
\cline{3-6}
& \multicolumn{1}{l}{}                                           & \multicolumn{1}{c}{{$\hat{\alpha}$}} & \multicolumn{1}{c}{{$\hat{\alpha}$ \textbf{error}}} & \multicolumn{1}{c}{{$\hat{\alpha}$}} & \multicolumn{1}{c}{{$\hat{\alpha}$ \textbf{error}}}  \endfirsthead 
\hline
				\multirow{3}{*}{$10^{3}$} & 0.5                       & 0.50652                                              & 1.30426\%                                                                      & 0.50652                                             & 1.30426\%                                                                      \\*
				& 1.5                       & 1.51956                                              & 1.30426\%                                                                      & 1.51956                                             & 1.30426\%                                                                      \\*
				& 2.2                       & 2.22869                                              & 1.30426\%                                                                      & 2.22869                                             & 1.30426\%                                                                      \\* 
				\hline
				\multirow{3}{*}{$10^{5}$} & 0.5                       & 0.49811                                              & 0.37806\%                                                                      & 0.49811                                             & 0.37806\%                                                                      \\*
				& 1.5                       & 1.49433                                              & 0.37806\%                                                                      & 1.49433                                             & 0.37806\%                                                                      \\*
				& 2.2                       & 2.19168                                              & 0.37806\%                                                                      & 2.19168                                             & 0.37806\%                                                                      \\
\end{longtable}
			\normalsize
			
			\small
			\begin{longtable}
{cccccc} 
	\hline
\multirow{2}{*}{Sample Size} & \multicolumn{1}{l}{\multirow{2}{*}{$\alpha$}} & \multicolumn{2}{c}{\pkg{ptsuite}}                                                                                                               & \multicolumn{2}{c}{\pkg{evir}}                                                                                                               \\* 
\cline{3-6}
& \multicolumn{1}{l}{}                                           & \multicolumn{1}{c}{{$\hat{\alpha}$}} & \multicolumn{1}{c}{{$\hat{\alpha}$ \textbf{error}}} & \multicolumn{1}{c}{{$\hat{\alpha}$}} & \multicolumn{1}{c}{{$\hat{\alpha}$ \textbf{error}}}  \endfirsthead 
\hline
				\multirow{3}{*}{$10^{3}$}    & 0.5                       & 0.50652                                  & 1.3043\%                                  & 0.51762                               & 3.5248\%                                \\*
				& 1.5                       & 1.51956                                  & 1.3043\%                                  & 1.62836                               & 8.5575\%                                \\*
				& 2.2                       & 2.22869                                  & 1.3043\%                                  & 2.45783                               & 11.7196\%                               \\* 
				\hline
				\multirow{3}{*}{$10^{5}$}  & 0.5                       & 0.49811                                  & 0.3781\%                                  & 0.50211                               & 0.4217\%                                \\*
				& 1.5                       & 1.49433                                  & 0.3781\%                                  & 1.55212                               & 3.4752\%                                \\*
				& 2.2                       & 2.19168                                  & 0.3781\%                                  & 2.30633                               & 4.8334\%                                \\* 
				\hline
				\multirow{3}{*}{$10^{6}$} & 0.5                       & 0.49993                                  & 0.0134\%                                  & 0.47815                               & 4.3709\%                                \\*
				& 1.5                       & 1.49980                                  & 0.0134\%                                  & 1.40399                               & 6.4005\%                                \\
\end{longtable}
			\normalsize
			
			\small
			\begin{longtable}
{cccccc} 
	\hline
\multirow{2}{*}{Sample Size} & \multicolumn{1}{l}{\multirow{2}{*}{$\alpha$}} & \multicolumn{2}{c}{\pkg{ptsuite}}                                                                                                               & \multicolumn{2}{c}{\pkg{DeMAND}}                                                                                                               \\* 
\cline{3-6}
& \multicolumn{1}{l}{}                                           & \multicolumn{1}{c}{{$\hat{\alpha}$}} & \multicolumn{1}{c}{{$\hat{\alpha}$ \textbf{error}}} & \multicolumn{1}{c}{{$\hat{\alpha}$}} & \multicolumn{1}{c}{{$\hat{\alpha}$ \textbf{error}}}  \endfirsthead 
\hline
				\multirow{3}{*}{$10^{2}$}   & 0.5                       & 0.574330                                     & 14.8661\%                                 & 1.608003                                    & 221.6007\%                                \\*
				& 1.5                       & 1.722991                                     & 14.8661\%                                 & 2.824010                                    & 88.2673\%                                 \\*
				& 2.2                       & 2.527054                                     & 14.8661\%                                 & 3.675215                                    & 67.0552\%                                 \\* 
				\hline
				\multirow{3}{*}{$10^{3}$}  & 0.5                       & 0.506521                                     & 1.3043\%                                  & 1.524971                                    & 204.9941\%                                \\*
				& 1.5                       & 1.519564                                     & 1.3043\%                                  & 2.574912                                    & 71.6608\%                                 \\*
				& 2.2                       & 2.228694                                     & 1.3043\%                                  & 3.309871                                    & 50.4487\%                                 \\* 
				\hline
				\multirow{3}{*}{$10^{4}$} & 0.5                       & 0.496634                                     & 0.6732\%                                  & 1.493813                                    & 198.7627\%                                \\*
				& 1.5                       & 1.489903                                     & 0.6732\%                                  & 2.481440                                    & 65.4293\%                                 \\
\end{longtable}
			\normalsize
			\small	
			\begin{longtable}
{cccccc} 
	\hline
\multirow{2}{*}{Sample Size} & \multicolumn{1}{l}{\multirow{2}{*}{$\alpha$}} & \multicolumn{2}{c}{\pkg{ptsuite}}                                                                                                               & \multicolumn{2}{c}{\pkg{laeken}}                                                                                                               \\* 
\cline{3-6}
& \multicolumn{1}{l}{}                                           & \multicolumn{1}{c}{{$\hat{\alpha}$}} & \multicolumn{1}{c}{{$\hat{\alpha}$ \textbf{error}}} & \multicolumn{1}{c}{{$\hat{\alpha}$}} & \multicolumn{1}{c}{{$\hat{\alpha}$ \textbf{error}}}  \endfirsthead 
\hline
				\multirow{3}{*}{$10^{3}$} & 1.5                       & 1.59742                                              & 6.49447\%                                                      & 1.5974                                              & 6.49362\%                                                      \\*
				& 2.2                       & 2.28164                                              & 3.71086\%                                                      & 2.40145                                             & 9.15664\%                                                      \\*
				& 5                         & 5.10218                                              & 2.04355\%                                                      & 6.06443                                             & 21.28859\%                                                     \\* 
				\hline
				\multirow{3}{*}{$10^{5}$} & 1.5                       & 1.49603                                              & 0.26447\%                                                      & 1.49015                                             & 0.65662\%                                                      \\*
				& 2.2                       & 2.19049                                              & 0.43230\%                                                      & 2.18272                                             & 0.78563\%                                                      \\*
				& 5                         & 4.97953                                              & 0.40947\%                                                      & 4.93508                                             & 1.29831\%                                                      \\
\end{longtable}
			\normalsize
			
			\small
			\begin{longtable}
{ccccccc} 
\hline
\multirow{2}{*}{Sample Size} &
\multicolumn{1}{c}{\multirow{2}{*}{\code{k}}} & \multicolumn{1}{c}{\multirow{2}{*}{$\alpha$}} & \multicolumn{2}{c}{\pkg{ptsuite}}                                                                                                               & \multicolumn{2}{c}{\pkg{laeken}}                                                                                                               \\* 
\cline{4-7}
& \multicolumn{2}{l}{}                                           & \multicolumn{1}{c}{{$\hat{\alpha}$}} & \multicolumn{1}{c}{{$\hat{\alpha}$ \textbf{error}}} & \multicolumn{1}{c}{{$\hat{\alpha}$}} & \multicolumn{1}{c}{{$\hat{\alpha}$ \textbf{error}}}  \endfirsthead 
\hline

				\multirow{3}{*}{$10^{3}$} & \multirow{3}{*}{900}     & 2.2                       & 2.2055                                               & 0.24996\%                                                      & 2.20475                                             & 0.21581\%                                                      \\*
				&                          & 1.5                       & 1.50375                                              & 0.24996\%                                                      & 1.50324                                             & 0.21581\%                                                      \\*
				&                          & 0.5                       & 0.50125                                              & 0.24996\%                                                      & 0.50108                                             & 0.21581\%                                                      \\* 
				\hline
				\multirow{3}{*}{$10^{5}$} & \multirow{3}{*}{99,900}  & 2.2                       & 2.19203                                              & 0.36248\%                                                      & 2.19201                                             & 0.36301\%                                                      \\*
				&                          & 1.5                       & 1.49456                                              & 0.36248\%                                                      & 1.49455                                             & 0.36301\%                                                      \\*
				&                          & 0.5                       & 0.49819                                              & 0.36248\%                                                      & 0.49818                                             & 0.36301\%                                                      \\
\end{longtable}
			\normalsize
			
		\end{footnotesize}
	\end{appendix}
	
	
\end{document}